\documentclass[usenatbib]{mn2e}
\usepackage{psfig}

\usepackage{epsfig}
\usepackage{amsmath}
\usepackage{lscape}
\usepackage{rotating}
\usepackage{hyperref}
\usepackage[labelsep=period,labelfont=bf]{caption}

\def\apj{ApJ}
\def\apjl{ApJL}
\def\mnras{MNRAS}
\def\pasp{PASP}

\def\araa{ARAA}
\def\aap{A\&A}

\def\apjs{ApJS}

\def\nat{Nature}

\def\gs{\mathrel{\raise0.35ex\hbox{$\scriptstyle >$}\kern-0.6em\lower0.40ex\hbox{{$\scriptstyle \sim$}}}} 
\def\ls{\mathrel{\raise0.35ex\hbox{$\scriptstyle <$}\kern-0.6em\lower0.40ex\hbox{{$\scriptstyle \sim$}}}}

\def\Wm2{\,\hbox{W}\,\hbox{m}^{-2}} 
\def\gsim{\mathrel{\raise0.35ex\hbox{$\scriptstyle >$}\kern-0.6em\lower0.40ex\hbox{{$\scriptstyle \sim$}}}} 
\def\lsim{\mathrel{\raise0.35ex\hbox{$\scriptstyle <$}\kern-0.6em\lower0.40ex\hbox{{$\scriptstyle \sim$}}}}
\newcommand{\appropto}{\mathrel{\vcenter{\offinterlineskip\halign{\hfil$##$\cr\propto\cr\noalign{\kern2pt}\sim\cr\noalign{\kern-2pt}}}}}

\begin{document}

\title[The Far-IR properties of ALMA SMGs]{An ALMA Survey of
  Sub-millimetre Galaxies in the Extended Chandra Deep Field South: The
  Far-Infrared Properties of SMGs}

\author[Swinbank et al.]
{\parbox[h]{\textwidth}
{A.\,M.\ Swinbank,$^{\,1,*}$
J.\, M.\ Simpson,$^{\,1}$
Ian Smail,$^{\,1}$
C.\, M.\ Harrison,$^{\, 1}$
J.\, A.\ Hodge,$^{\,2}$
A.\ Karim,$^{\,3}$
F.\ Walter,$^{\,2}$
D.\,M.\ Alexander,$^{\, 1}$
W.\, N.\ Brandt,$^{\,4}$
C.\, de Breuck,$^{\,5}$
E.\, da Cunha,$^{\,2}$
S.\,C.\ Chapman,$^{\,6}$
K.\,E.\,K\ Coppin,$^{\,7}$
A.\,L.\,R.\ Danielson,$^{\,1}$
H.\ Dannerbauer,$^{8}$
R.\, Decarli,$^{\,2}$
T.\,R.\ Greve,$^{\,9}$
R.\,J.\ Ivison,$^{\,10}$
K.\, K.\ Knudsen,$^{\,11}$
C.\, D.\, P.\ Lagos,$^{\,5}$
E.\ Schinnerer,$^{\,2}$
A.\,P.\ Thomson,$^{\,1}$
J.\,L.\ Wardlow,$^{12}$
A.\ Wei\ss$^{\,13}$
\& P.\ van der Werf$^{\,14}$
}
\vspace*{4pt} \\ 
$^1$Institute for Computational Cosmology, Durham University, South Road, Durham, DH1 3LE, UK\\
$^2$Max-Planck-Institut f\"ur Astronomie, K\"onigstuhl 17, D-69117 Heidelberg, Germany\\
$^3$Argelander-Institute of Astronomy, Bonn University, Auf dem Huegel 71, D-53121 Bonn, Germany \\
$^4$Department of Astronomy \& Astrophysics, 525 Davey Lab, The Pennsylvania State University, University Park, PA 16802, USA\\
$^5$European Southern Observatory, Karl-Schwarzschild Strasse 2, D-85748 Garching, Germany\\
$^6$Dalhousie University, Halifax, Nova Scotia B3H 4R2, Canada\\
$^7$Centre for Astrophysics Research, University of Hertfordshire, College Lane, Hatfield, Herts AL10 9AB\\
$^8$Universit\"at Wien, Institut f\"ur Astrophysik,  T\"urkenschanzstra\ss e 17, 1180 Wien, Austria\\
$^{9}$University College London, Department of Physics \& Astronomy, Gower Street, London, WC1E 6BT, UK \\
$^{10}$Institute for Astronomy, University of Edinburgh, Blackford Hill, Edinburgh EH9 3HJ\\
$^{11}$Department of Earth and Space Sciences, Chalmers University of Technology, Onsala Space Observatory, SE-43992 Onsala, Sweden\\
$^{12}$Dark Cosmology Centre, Niels Bohr Institute, University of Copenhagen, Denmark\\
$^{13}$Max-Planck-Institut f\"ur Radioastronomie, Auf dem H\"ugel 69, D-53121 Bonn, Germany\\
$^{14}$Leiden Observatory, Leiden University, PO Box 9513, 2300 RA Leiden, Netherlands\\
$^{*}$Email: a.m.swinbank@dur.ac.uk\\
\vspace{-1cm}
}

\maketitle

\begin{abstract}
  We exploit ALMA 870-$\mu$m (345\,GHz) observations of sub-millimetre
  sources in the Extended {\it Chandra} Deep Field South to investigate
  the far-infrared properties of high-redshift sub-millimetre galaxies
  (SMGs).  Using the precisely located 870\,$\mu$m ALMA positions of 99
  SMGs, together with 24$\mu$m and radio imaging of this field, we
  deblend the \emph{Herschel}\,/\,SPIRE imaging of this region to
  extract their far-infrared fluxes and colours.  The median redshifts
  for ALMA LESS (ALESS) SMGs which are detected in at least two SPIRE
  bands increases as expected with wavelength of the peak in their
  SEDs, with $z$\,=\,2.3\,$\pm$\,0.2, 2.5\,$\pm$\,0.3 and
  3.5\,$\pm$\,0.5 for the 250, 350 and 500-$\mu$m peakers respectively.
  We find that 34 ALESS SMGs do not have a $>$3\,$\sigma$ counterpart
  at 250, 350 or 500$\mu$m.  These galaxies have a median photometric
  redshift derived from the rest-frame UV--mid-infrared SEDs of
  $z$\,=\,3.3\,$\pm$\,0.5, which is higher than the full ALESS SMG
  sample; $z$\,=\,2.5\,$\pm$\,0.2.  Using the photometric redshifts
  together with the 250--870$\mu$m photometry, we estimate the
  far-infrared luminosities and characteristic dust temperature of each
  SMG.  The median infrared luminosity and characteristic dust
  temperature of the $S_{\rm 870\mu m}> $\,2\,mJy SMGs is $L_{\rm
    IR}$\,=\,(3.0\,$\pm$\,0.3)\,$\times$\,10$^{12}$\,$L_{\odot}$ (star
  formation rate of SFR\,=\,300\,$\pm$\,30\,$M_{\odot}$\,yr$^{-1}$) and
  $T_{\rm d}$\,=\,32\,$\pm$\,1\,K ($\lambda_{\rm
    peak}$\,=\,93\,$\pm$\,4\,$\mu$m).  At a fixed luminosity, the
  characteristic dust temperature of these high-redshift SMGs is
  $\Delta T_{\rm d}$\,=\,3--5\,K lower than comparably luminous
  galaxies at $z$\,=\,0, reflecting the more extended star formation
  occurring in these systems.  By extrapolating the 870$\mu$m number
  counts to $S_{\rm 870}$\,=\,1\,mJy, we show that the contribution of
  $S_{\rm 870\mu m}\geq $\,1\,mJy SMGs to the cosmic star formation
  budget is 20\% of the total over the redshift range $z\sim $\,1--4.
  We derive a median dust mass for these galaxies of $M_{\rm
    d}$\,=\,(3.6\,$\pm$\,0.3)\,$\times$\,10$^{8}$\,M$_{\odot}$ and by
  adopting an appropriate gas-to-dust ratio, we estimate that the
  typical molecular mass of the ALESS SMGs in our sample is $M_{\rm
    H_2}$\,=\,(4.2\,$\pm$\,0.4)\,$\times$\,10$^{10}$\,M$_{\odot}$.
  Together with the average stellar masses of SMGs, $M_{\rm
    \star}$\,=\,(8\,$\pm$\,1)\,$\times$\,10$^{10}$\,M$_{\odot}$, this
  suggests an average molecular gas fraction of $\sim $\,40\%.
  Finally, we use our estimates of the H$_2$ masses to show that SMGs
  with $S_{\rm 870\mu m}> $\,1\,mJy ($L_{\rm IR}\gsim
  $\,10$^{12}$\,L$_{\odot}$) contain $\sim $\,10\% of the $z\sim $\,2
  volume-averaged H$_2$ mass density at this epoch.
\end{abstract}

\begin{keywords}
  galaxies: starburst, galaxies: evolution, galaxies: high-redshift
\end{keywords}

\section{Introduction}

Ultra-luminous Infrared Galaxies (ULIRGs; \citealt{Sanders96}) have
total infrared luminosities $> $\,10$^{12-13}$\,L$_{\odot}$ and implied
star formation rates (SFR) $>$\,100--1000\,M$_{\odot}$\,yr$^{-1}$.
Their low space densities mean that ULIRGs contribute $\ll $\,1\% of
the volume average star formation density at $z$\,=\,0.  However, the
first deep, single-dish bolometer surveys in the 870-$\mu$m atmospheric
window uncovered high number densities of high-redshift sub-mm galaxies
(SMGs) at mJy flux density levels \citep{Smail97,Hughes98,Barger98}.
Subsequent spectroscopic studies of the radio-\,/\,mid-infrared
detected subset of the population gave a median redshift of $z\sim
$\,2.5 \citep{Chapman05a}, confirming their ULIRG-like luminosities
\citep{Kovacs06,Coppin08a,Magnelli12} and demonstrated that ULIRGs
undergo a 1000-fold increase in space density from $z$\,=\,0 to $z\sim
$\,2.5.  Thus, in contrast to the local Universe, ULIRGs are a
non-negligible component of the star-forming population at $z\sim
$\,2--3
\citep[e.g.\ ][]{Hughes98,Lilly99,Blain99a,Chapman05a,Wardlow11,Barger13,Casey13}.

These 870-$\mu$m-selected samples remain the best-studied SMGs, and
links have been proposed between SMGs, QSOs and the formation phase of
massive galaxies at high redshift
\citep[e.g.\ ][]{Genzel03,Swinbank06b,Coppin08b,Daddi09,Hickox12}.
SMGs are therefore a potentially important element in models of galaxy
formation.  However, these evolutionary links are still unproven,
although it is clear that most luminous SMGs lie above the purported
``main sequence'' of star-forming galaxies in the star formation
rate--M$_{\star}$ plane
\citep[e.g.\ ][]{Daddi07,Noeske07,Rodighiero11,Wardlow11}.  Along with
corroborating kinematic and morphological evidence
\citep{Tacconi06,Tacconi08,Engel10,Swinbank10b,SAZ12}, inevitably it
has been argued that SMGs can be understood as ``scaled-up'' analogs of
local ULIRGs which are dominated by merger induced starbursts.
It has also been proposed that lower luminosity SMGs overlap
with the ``main sequence'', leading to claims of an apparent dichotomy
within this population: i.e.\ between a true ``starburst'' population
and slightly less active galaxies, potentially driven by secular
processes \citep{Rodighiero11}.

Theoretical attempts to reproduce basic properties of SMGs,
such as 870-$\mu$m number counts, have also led to an equally wide variety
of conclusions about the nature and diversity of this population. For
example, early models from \citet{Baugh05} and \citet{Granato06}
include two recipes for star formation; ``burst'' and ``quiescent'',
with SMGs corresponding to the most extreme starburst systems.  One key
strength of these models is that they are also required to fit the
galaxy population at $z$\,=\,0, which forces the former to adopt a
burst initial mass function (IMF) biased to high-mass stars (allowing
relatively low-mass galaxies to produce intense starbursts;
\citealt{Baugh05}).  In contrast, numerical models of \citet{Hayward11}
(see also \citealt{Narayanan09} and \citealt{Dave10}) suggest that SMGs
are dominated by secular bursts in gas-rich disk galaxies, with
``standard'' IMFs.  Hence theoretical models variously predict the SMG
population to comprise low-mass merging starbursts (with unusually low
mass-to-light ratios; e.g.\ \citealt{Baugh05}) and isolated, gas-rich
disk galaxies undergoing secular bursts
\citep[e.g.\ ][]{Dave10,Hayward12}.

This rich array of theoretical options demonstrates the power of SMGs
to distinguish between the competing galaxy evolution theories.
Observations of the most basic properties of SMGs, such as their
redshift distribution, bolometric luminosities, cold molecular gas
dynamics and gas mass, and contribution to the cosmic star formation
density should have the power to distinguish between the theoretical
models.

However, the coarse resolution of single dish sub-mm telescopes
(typically $\sim$\,15--20$''$) means that identifying the SMG
counterparts has to date, relied on correlations between their
sub-millimetre emission and that in other wave-bands where higher
spatial resolution is available (usually the radio and\,/\,or
mid-infrared; e.g.\ \citealt{Ivison02,Ivison05,Ivison07,Pope06}).  The
spectral energy distributions (SEDs) in these other wave-bands have
positive K-corrections, making counterparts which either lie at higher
redshift or that have colder-than-average dust temperatures impossible
to identify.  Indeed, in sub-millimetre surveys typically 40--50\% of
sub-mm sources lack ``robust'' counterparts in the radio or mid-infrared
(e.g.\ \citealt{Biggs11}, see also \citealt{Lindner11}) and it is
unknown whether the unidentified SMGs have the same redshift
distribution (or are representative) of the radio-identified subset,
potentially biasing the current observational results.

To circumvent the problem of missing- (and mis-) identifications and so
characterize the {\it whole} population of bright SMGs in an unbiased
manner requires precisely locating the sub-mm emission using sub/mm
interferometers \citep[e.g.\ ][]{Wang11,Smolcic12}.  Recently, we have
undertaken an ALMA survey of the 126 sub-mm sources in the
0.5\,$\times$\,0.5 degree Extended {\it Chandra} Deep Field South
(ECDFS), taken from the ``LESS'' survey \citep{Weiss09}.  The ALMA data
precisely locate the SMGs, {\it directly} pin-pointing the source(s)
responsible for the sub-mm emission (to within $<$\,0.3$''$), without
recourse to statistical radio\,/\,mid-IR associations and so yielding
unambiguous identifications for the majority of the SMGs.  The first
results from our survey include the source catalog and multiplicity of
bright SMGs \citep{Hodge13}; the first high-resolution sub-mm counts
\citep{Karim13}; optical and near-infrared properties, photometric
redshift distribution and stellar masses (Simpson et al.\ 2013);
serendipitous identification [C{\sc ii}] at $z$\,=\,4.4 in two SMGs
\citep{Swinbank12c} and subsequent $^{12}$CO follow-up in these
galaxies \citep{Huynh13}; X-ray properties and AGN fraction of ALMA
SMGs \citep{Wang13}; far-infrared--radio correlation of SMGs (Thomson
et al. 2014 submitted) and \emph{HST} morphologies (Simpson et
al.\ 2014 in prep).

In this paper, we extend these analysis and exploit the \emph{Herschel}
Space Observatory PACS and SPIRE imaging of the ECDFS to derive the
far-infrared/sub-millimeter properties of the SMGs pinpointed by ALMA.
In particular, we use the far-infrared (70--870$\mu$m) and 1.4\,GHz
radio photometry together with new optical/mid-infrared-derived
photometric redshifts from Simpson et al.\ (2013) to derive the
far-infrared luminosities, characteristic dust temperatures, dust
masses and investigate their evolution with redshift.  In \S2 we
describe the observations and deblending techniques; in \S3 we present
the far-infrared properties (colours, luminosities, dust masses) of
both individual ALESS SMGs and those of the stacks of subsets of
population.  We present our main conclusions in \S4.  Throughout the
paper, we adopt a Chabrier IMF \citep{Chabrier03} and use a
$\Lambda$CDM cosmology \citep{Spergel07} with
$\Omega_{\Lambda}$\,=\,0.73, $\Omega_{\rm m}$\,=\,0.27, and $H_{\rm
  0}$\,=\,72\,km\,s$^{-1}$\,Mpc$^{-1}$.

%
%
\begin{figure*}
  \centerline{\psfig{file=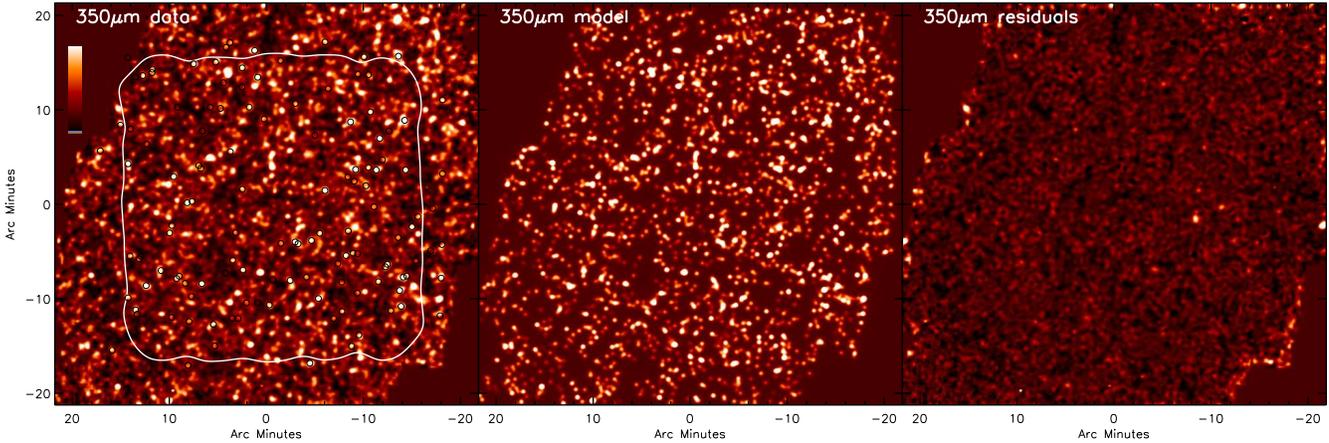,angle=0,width=7in}}
\caption{{\it Left:} 350$\mu$m\,/\,SPIRE image of ECDFS.  We mark the
  position of the LESS SMGs from \citet{Weiss09}.  The contour denotes
  the area enclosed by the 1.2\,mJy r.m.s. of the LABOCA survey
  \citep{Weiss09}; {\it Center:} Best-fit 350$\mu$m model image of the
  field from our deblending algorithm which uses the MIPS\,24$\mu$m,
  radio and ALMA positions as a positional prior catalog; {\it Right:}
  350$\mu$m residuals between (after subtracting the best-fit model
  from the data).  All panels are centered at $\alpha:$\,03\,32\,32.25,
  $\delta:$\,$-$27:48:17.2 (J2000) with North up and East Left.
}
\label{fig:ECDFS350}
\end{figure*}

%
%
\begin{table}
{\footnotesize
\begin{center}
\caption{Stacked far-infrared photometry for ALMA SMGs}
\begin{tabular}{lccc}
\hline
\hline
                             & ALL                     & S$_{\rm 1.4\,GHz}$             & S$_{\rm 1.4\,GHz}$ \\
                             &                         & $> $\,25$\mu$Jy              & $< $\,25$\mu$Jy \\
\hline
N                            & 99                      & 46                           & 53                            \\
\hline                 
S$_{\rm 70\mu m}$\,(mJy)       & $<$0.45  (0.1)          & 1.3\,$\pm\,$0.4 (0.1)        & $<$\,1.3 (0.1)            \\
S$_{\rm 100\mu m}$\,(mJy)      & 3.0\,$\pm$\,1.0 (0.1)   & 3.9\,$\pm$\,1.4 (0.2)        & $<$\,2.5 (0.2)            \\
S$_{\rm 160\mu m}$\,(mJy)      & 9.7\,$\pm$\,1.4 (0.3)   & 11.7\,$\pm$\,1.9 (0.4)       & 8.0\,$\,\pm$\,1.8 (0.4)  \\
S$_{\rm 250\mu m}$\,(mJy)      & 16.0\,$\pm$\,1.1 (0.4)  & 19.3\,$\pm$\,1.5 (0.7)       & 13.6\,$\pm$\,1.5 (0.5)    \\
S$_{\rm 350\mu m}$\,(mJy)      & 20.6\,$\pm$\,1.2 (0.5)  & 23.5\,$\pm$\,2.2 (0.8)       & 19.2\,$\pm$\,1.5 (0.7)    \\
S$_{\rm 500\mu m}$\,(mJy)      & 18.5\,$\pm$\,1.1 (0.8)  & 20.4\,$\pm$\,1.2 (0.9)       & 16.1\,$\pm$\,1.3 (0.8)    \\
S$_{\rm 870\mu m}$\,(mJy)      & 4.0\,$\pm$\,0.3 (0.4)   & 4.5\,$\pm$\,0.31 (0.4)       & 3.6\,$\pm$\,0.37 (0.4)    \\
S$_{\rm 1.4GHz}$\,($\mu$Jy)    & 13.8\,$\pm$\,2.4 (0.9)  & 28.9\,$\pm$\,3.9 (1.3)       & 4.1\,$\pm$\,2.4 (1.4)      \\
\hline
\label{table:stack}
\end{tabular}
\end{center}
\noindent{\footnotesize Notes: The errors are those on the bootstrap of
  the distribution.  The error in parenthesis denoted the noise in the
  stacked map.}
}
\end{table}

%
%
\section{Observations \& Analysis}

\subsection{ALMA}

Details of the ALMA observations of the sub-millimeter sources from the
ALMA LESS (ALESS) survey are described in \citet{Hodge13} \citep[see
  also][]{Karim13}.  Briefly, observations of 122 of the 126 LESS
sources were obtained with ALMA in Cycle 0 at 345\,GHz (Band 7) with a
dual polarisation setup in the compact configuration (yielding a
synthesised beam of $\sim$\,1.6$''\times$1.2$''$).  The ALMA primary
beam, 17.3$''$ FWHM at our observing frequency, is sufficient to
encompass the error-circles of the sub-millimeter sources from the LESS
maps, $\lsim$\,5$''$ \citep{Weiss09}.  The observations employed 12--15
antennae and were obtained between 2011 October and 2011 November in
good conditions, PWV\,$\lsim$\,0.5\,mm.  Phase and bandpass calibration
was based on J0403$-$360, J0538$-$440 respectively and flux calibration
performed on available planets at the time of observation.  The data
were processed with the Common Astronomy Software Application ({\sc
  casa}; \citealt{McMullin07}).  The resulting velocity integrated
continuum maps reach typical noise levels of
$\sigma$\,=\,0.4--0.5\,mJy\,beam$^{-1}$, a factor $\sim $\,3\,$\times$
more sensitive than the original LABOCA discovery map and, critically,
with a beam that is $\sim $\,200\,$\times$ smaller in area than that of
LABOCA.

From the ALMA maps, \citet{Hodge13} extract 99\,SMGs with S\,/\,N$>$3.5
from the best maps (noise $<$\,0.6\,mJy and synthesised beam with axial
ratio $<$\,2).  This selection provides an acceptable trade off between
source reliability and spurious sources.  Indeed, using the background
fluctuations in the map, \citet{Karim13} demonstrate that we expect
only one SMG in the catalog to be spurious and one SMG to be missed.
We note that \citet{Hodge13} also provide a catalog of a further 32
ALMA SMGs which are from either the shallower maps (noise levels
$\sigma<$\,0.6--1.0\,mJy), or $>$\,4\,$\sigma$ sources within
2\,$\times$ the primary beam of the best maps, but these
``supplementary'' catalogs are expected to have lower completeness and
a higher spurious fraction and so conservatively we do not use these
sources in our analysis here.

\begin{figure*}
  \centerline{\psfig{file=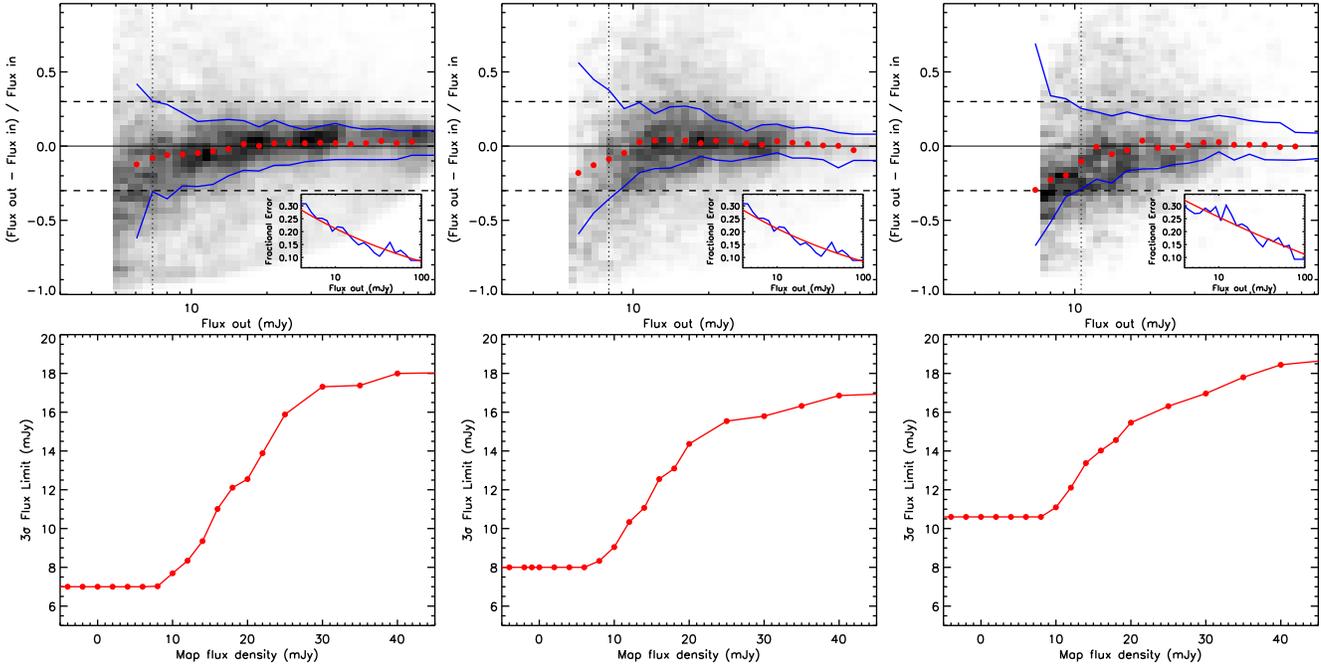,angle=0,width=7in}}
\caption{{\it Top:} Error distributions for the deblended SPIRE maps
  based on the prior catalogs.  These show the results from simulations
  for extraction of injected sources into the maps.  The solid
  curves enclose 60\% of the points at each flux, and we define the
  detection limit as the flux density where 68\% of the injected
  galaxies are recovered with a flux density error $<$\,30\% (shown by
  the horizontal dashed lines).
  We derive detection limits at 250, 350, and 500$\mu$m of 7.0, 8.0 and
  10.6\,mJy respectively, as shown by the vertical dotted lines.  The
  insets show the fractional flux density error as a function of
  recovered flux density (and a low order polynomial fit) of the
  injected sources.  These flux density fractional errors are combined
  with the Monte-Carlo errors from the deblending to derive errors on
  the flux density of each galaxy.  {\it Bottom:} For galaxies which
  are not detected in the deblended map (either because they are too
  faint, or they lie in the halo of a nearby ``bright'' source which
  effectively increases the local background), we derive their
  3-$\sigma$ upper limit by using the recovered flux density for all
  injected sources as a function of the local background.  These limits
  are shown for 250, 350 and 500$\mu$m in the lower panels.}
\label{fig:DeblendErrors}
\end{figure*}

\subsection{1.4\,GHz VLA Radio \& 24\,$\mu$m {\it Spitzer} Imaging}

The ECDFS was observed with the JVLA in D, C, B and A configurations in
2007 June to September, reaching an r.m.s. of $\sigma_{\rm
  1.4\,GHz}$\,=\,8.3\,$\mu$Jy with a synthesised beam of
2.8$''$\,$\times$\,1.6$''$.  These data and catalogs are described in
detail in \citet{Miller13} \citep[see also][]{Biggs11}.

{\it Spitzer}\,/\,MIPS 24\,$\mu$m imaging is also available for the ECDFS
as part of the Far-Infrared Deep Extra-galactic Legacy (FIDEL) survey.
This MIPS 24\,$\mu$m imaging provides an important addition in the
construction of a positional prior catalog which is used to deblend the
\emph{Herschel} images (see \S~\ref{sec:deblending}) as well as
providing a constraint used in the far-infrared SEDs of the ALMA SMGs.
We obtained the reduced MIPS 24\,$\mu$m images of the ECDFS from the
NASA Infrared Astronomy Archive \footnote[7]{\tt
  http://irsa.ipac.caltech.edu/data/SPITZER/FIDEL/}.  This imaging
covers the entire ECDFS survey area, and we extract a catalog of $\sim
$\,3600 sources in the ECDFS down to a 5-$\sigma$ depth of
$\sim$\,56\,$\mu$Jy (aperture corrected).

\subsection{Herschel\,/\,PACS and SPIRE Imaging}

\emph{Herschel}\,/\,PACS observations covering the ECDFS at 100 and
160\,$\mu$m and the CDFS at 70, 100 and 160\,$\mu$m were taken as part
of the PACS Evolutionary Probe (PEP) survey (the CDFS lies in the
central 0.11-square degrees of the ECDFS).  These data and deblended
catalogs are described in \citet{Lutz11} and \citet{Magnelli13} and
reach 1-$\sigma$ sensitivities of 0.2--0.4\,mJy (CDFS) and 1--2.6\,mJy
(ECDFS).  We match our ALESS catalog with those from \citet{Magnelli13}
(with a matching radius of 1.5$''$) and include the 70--160\,$\mu$m
photometry for the ALESS SMGs from the CDFS and the ECDFS wherever
possible in our analysis below.

\emph{Herschel}\,/\,SPIRE 250, 350 and 500$\mu$m observations covering
ECDFS were taken as part of the \emph{Herschel} Multi-tiered
Extra-galactic Survey (HerMES) guaranteed time program (as described in
\citealt{Oliver12}).  In total, ECDFS was observed for 32.4\,ks at 250,
350 and 500\,$\mu$m in $\sim$\,1.8\,ks blocks.  For each observation,
we retrieved the Level~2 data product from the \emph{Herschel} ESA
archive and aligned and co-added the maps.  The final combined maps
reach a 1-$\sigma$ noise level of 1.6, 1.3 and 1.9\,mJy at 250, 350
and 500\,$\mu$m respectively \citep[see][for a detailed
  description of the observations]{Oliver12}.  

To align the SPIRE maps to the ALMA and radio astrometry, we aligned
the SPIRE maps at 250, 350 and 500$\mu$m on the VLA radio positions,
identifying and applying shifts of $\Delta <$\,1.5$''$ in all cases.

%
%
\begin{figure*}
  \centerline{\psfig{file=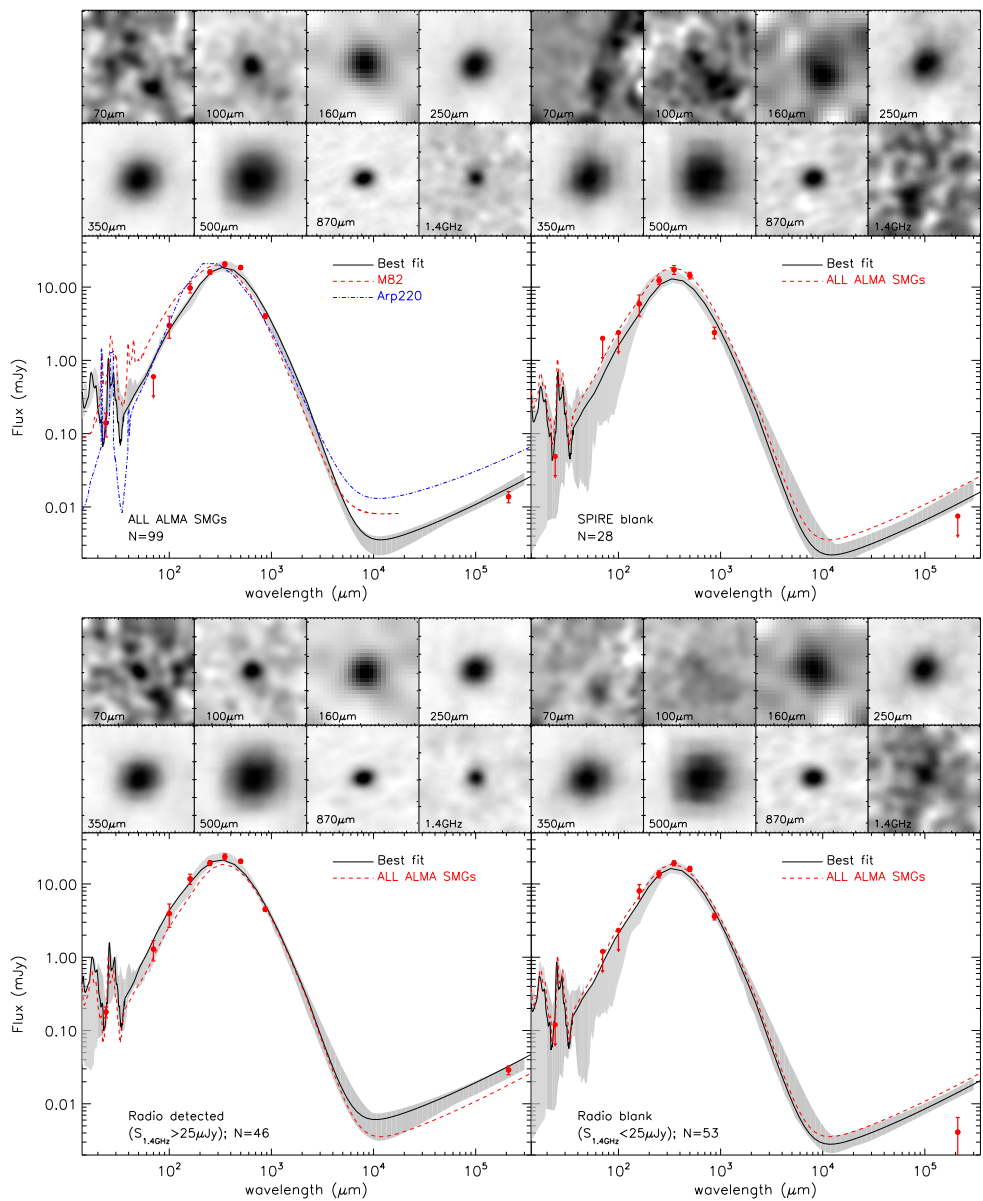,angle=0,width=5in}}
\caption{{\it Top Left:} Stacked (observed frame) spectral energy
  distribution for all 99 ALESS SMGs in our sample.  The solid curve
  shows the best-fit SED from our template library (the best-fit
  template has a redshift of $z$\,=\,1.8), with the shaded region shows
  the range of acceptable solutions.  We also overlay the SEDs of M\,82
  and Arp\,220 (redshifted to $z$\,=\,2.5 and normalised to the peak).
  {\it Top Right:} The observed composite SED for the 34 individually
  SPIRE undetected ALESS SMGs.  The solid curve denotes the best-fit
  template (which has a best-fit of $z$\,=\,3.5) and the dashed curve
  shows the best-fit SED to the ``All'' composite in the left hand
  panel.  This ``SPIRE-undetected'' stack appears to have the same
  far-infrared colours as the ``All'' stack, but is lower luminosity at
  all wavelengths.  {\it Bottom Left:} Observed frame composite SED for
  the 46 individually radio-identified ALESS SMGs with the best-fit
  (solid) and ``All'' (dashed) composites overlaid.  In this plot, the
  best-fit template is for $z$\,=\,1.3. {\it Bottom Right:} The
  observed SED for the 53 ALESS SMGs which are (individually)
  non-detected at 1.4\,GHz, again with the best-fit (solid) and ``All''
  (dashed) composites overlaid.  The best-fit template is for
  $z$\,=\,3.5.  This ``radio-detected'' composite SED appears to have
  similar 250, 350 and 500-$\mu$m colours as the ``radio non-detected''
  composite, although the ``radio-detected'' composite has more flux at
  shorter (70--160-$\mu$m) wavelengths, most likely reflecting the
  differences in the photometric redshift distributions of the two
  sub-samples (assuming a fixed dust temperature).  For each sample, we
  also show the thumb-nail images from the stacks of the
  \emph{Herschel}\,/\,PACS\,+\,SPIRE (70--500$\mu$m), ALMA 870$\mu$m
  and VLA (1.4\,GHz) radio.
  In each of these thumbnail the major tick-marks are spaced by 10$''$.
}
\label{fig:HSOstack}
\end{figure*}

\subsubsection{Deblending SPIRE maps}
\label{sec:deblending}

Owing to the coarse beam size in the SPIRE maps, to measure reliable
far-infrared flux densities for individual galaxies we need to deblend
the SPIRE photometry for the effects of confusion.  We therefore
exploit the extensive multi-wavelength imaging of ECDFS to construct a
catalog of infrared- and radio-bright galaxies which can be used as
positional priors to deblend the SPIRE maps.  First, we combine the
$>$\,5-$\sigma$ MIPS 24\,$\mu$m and radio catalogs, removing any
sources within 1.5\,$''$ as duplicates (in this case, we remove the
lower signal-to-noise of the pair).  The 1.5$''$ offset we apply
represents an acceptable trade off between the resolution of the radio
map and centering precision of the 24\,$\mu$m data for a 5-$\sigma$
source.  We then add the positions of the 99 ALMA SMGs to this catalog,
again matching any 24\,$\mu$m or radio sources which lie within
1.5\,$''$ (the approximate resolution of the ALMA data) of the ALESS
SMG as the same ID.

To deblend the SPIRE map we develop a Monte Carlo algorithm.  At any
given position in the field, we extract a thumbnail which has an extent
$\pm$\,2.5\,$\times$ the FWHM of the beam at that wavelength.  We then
generate a new (blank) image and for each galaxy in the positional
prior catalog which lies within this area assign a random flux
densities (which lies between between zero and 1.3\,$\times$ the
maximum flux density of galaxies within the thumbnail).  We then
convolve this image with the relevant SPIRE point spread function (PSF)
and record the amplitude of the residuals and $\chi^2$.  From an
initial set of 1000 models, we identify the best-fit model (lowest
$\chi^2$).  We then repeat this process, creating a new set of images
with perturbed flux densities for each galaxy according to the FWHM of
the flux density distribution from the previous set of models.
This process repeats until all of the models in a given iteration are
within a $\Delta\chi^2$\,=\,1$\sigma$ of the best-fit.  Throughout this
process, we record the flux density distribution and $\chi^2$ of every
model attempted.

To ensure we do not ``over deblend'' the maps at longer wavelengths
where the PSF is larger, we follow \citet{Elbaz11} and when deblending
the 350$\mu$m image, we only include sources detected at $>
$\,2\,$\sigma$ at 250$\mu$m as positional priors in the 350$\mu$m image (and
similarly, for the 500$\mu$m we only include sources $>$2\,$\sigma$ at
350$\mu$m).

Before constructing a catalog, we must estimate both flux density
errors and upper limits for non-detections.  First, we determine the
detection limit by attempting to recover fake point sources which have
been randomly injected into the map (and positional prior catalog).  In
total, we inject $\sim $\,10,000 point sources into each of the 250,
350 and 500$\mu$m maps (one at a time) with fluxes between 0.5 and
100\,mJy and record the input ($F_{\rm in}$) and recovered ($F_{\rm
  out}$) flux density of the galaxy after deblending.  In
Fig.~\ref{fig:DeblendErrors} we plot the fractional flux density error
(($F_{\rm out}$-$F_{\rm in}$)/$F_{\rm in}$) as a function of $F_{\rm
  out}$ at each wavelength and contour the central 68\% of the
distribution.  Following \citet{Magnelli13}, we define the 3-$\sigma$
detection limit when 68\% of the distribution are recovered with a
fractional error less than 30\%.  In the ECDFS we derive 3-$\sigma$
detection limits of 7.0, 8.0 and 10.6\,mJy at at 250, 350 and 500$\mu$m
respectively.  These are similar to the faintest fluxes reported
galaxies in the ECDFS using the XID deblending procedure by
\citet{Roseboom10} (who derive fluxes for their faintest galaxies 6.5,
8.5 and 8.0\,mJy at 250, 350 and 500-$\mu$m respectively;
\citealt{Casey12}).  For galaxies which are detected above 3-$\sigma$,
we also calculate the fractional flux density error according to the
distributions shown in Fig.~\ref{fig:DeblendErrors} and to be
conservative, add this in quadrature to the errors derived from the
family of acceptable models from the Monte-Carlo deblending.

For the galaxies which have flux densities below these limits, we
calculate an upper limit.  This upper limit depends on its location in
the map.  For example, a source that lies within the beam of a nearby,
brighter source is more difficult to ``detect'' than an isolated source
since the effective background has increased due to the emission from
the nearby source and large PSF.  We therefore measure the recovered
flux density for all injected sources as a function of the local
background in the map.  Again, following \citet{Magnelli13}, we derive
the upper limit by identifying where 80\% of the injected sources have
a recovered flux density within 50\% of the input flux density.  We
show this distribution in Fig.~\ref{fig:DeblendErrors} and use this
distribution to assign upper limits for non-detections.  

To validate the deblending (and errors), we simulate a set of SPIRE
images using the SPIRE number counts and redshift distribution from
\citet{Clements12}.  We construct images at 250, 350 and 500\,$\mu$m
over a 1-square degree region (and include sources down to flux density
limits of 0.5\,mJy) and convolve the map with the relevant PSF.  We
note that we have not included any clustering of the sources in this
simple analysis.  To this image we then add Gaussian noise at the same
level as the SPIRE observations of the ECDFS.  To construct the
positional prior distribution in a comparable way to our ECDFS data, we
predict the 24\,$\mu$m flux density for each source injected into the
fake map using its redshift and 250\,$\mu$m flux density and using an
far-infrared SED randomly selected from the templates of
\citet{CharyElbaz01}.  We then construct a positional prior catalog for
all sources brighter than $S_{\rm 24\mu m}$\,=\,50\,$\mu$Jy and attempt
to recover the sources using our algorithm.  Using this method, the
ratio of the input flux density to that recovered at 250, 350 and
500\,$\mu$m is $S_{\rm \lambda}^{\rm in}$\,/\,$S_{\rm \lambda}^{\rm
  out}$\,=\,0.96\,$\pm$\,0.02, 0.97\,$\pm$\,0.02 and 1.17\,$\pm$\,0.12
respectively.

Having validated our approach, next we apply this algorithm to the
ECDFS at 250, 350 and 500$\mu$m, running the code in a grid of
(overlapping) regions in ECDFS, each of extent $\sim $\,5\,beams at the
relevant wavelength.  In Fig.~\ref{fig:ECDFS350} we show the 350$\mu$m
SPIRE map, with the positions of the LESS SMGs from \citealt{Weiss09}
highlighted, as well as the best-fit 350$\mu$m model of the field and
the residuals \footnotemark. \footnotetext{The SPIRE 250, 350 and
  500$\mu$m images, best fit models, residual maps, and deblended
  catalog for all galaxies in our prior catalog in the ECDFS (as well
  as the COSMOS, UDS and GOODS-North regions which we also use to
  validate our fluxes and errors compared to \citealt{Casey12}) are
  available at:
  \href{http://astro.dur.ac.uk/$\sim$ams/HSOdeblend/ECDFS/}{http://astro.dur.ac.uk/$\sim$ams/HSODeblend/}}

From the deblended catalog, we then extract the SPIRE photometry for
the ALESS SMGs and give these in Table~A1.  In
Fig.~\ref{fig:DataModelResiduals} we show example thumb-nails around
four ALESS SMG in our sample, the best-fit models and residuals at 250,
350 and 500$\mu$m (these four galaxies are randomly selected to span
the full range of 870$\mu$m flux density from LESS; \citealt{Weiss09}).
In each panel we show the positions of all of the galaxies in the
``prior'' catalog at that wavelength and also highlight the positions
of the ALESS SMGs.  The far infrared photometry of the ALESS SMGs (from
24$\mu$m to 1.4\,GHz) is given in Table~A1.  In cases where no flux is
detected, we quote 3-$\sigma$ upper limits.

As a final check, we compare the flux densities derived for the
galaxies in our prior catalog with those recently published by
\citet{Casey12} (which are based on the ``XID'' deblending procedures
from \citealt{Roseboom10}).  \citet{Casey12} report deblended 250, 350
and 500$\mu$m flux densities and spectroscopic redshifts for a sample
of $\sim $\,750 24$\mu$m- and 1.4\,GHz-detected galaxies with SPIRE
counterparts in the COSMOS, ECDFS, GOODS-N and Lockman Hole regions.
We cross correlate our deblended catalog with the \citet{Casey12}
catalog (including our own deblended maps of the COSMOS, ECDFS and
GOODS-N fields to improve the number of matches between samples) and
derive comparable flux densities at all three SPIRE wavelengths, with
$(F_{\rm Db}^\lambda$\,-\,$F_{\rm XID}^{\lambda}$)\,/\,$F_{\rm
  Db}^{\lambda}$\,=\,$-$0.03\,$\pm$\,0.02, 0.08\,$\pm$\,0.04 and
0.08\,$\pm$\,0.10 at $\lambda$\,=\,250, 350 and 500$\mu$m respectively.
This suggests that the ``XID'' and our deblending produce consistent
flux densities to $\lsim$\,10\% accuracy.

\begin{figure*}
  \centerline{
    \psfig{file=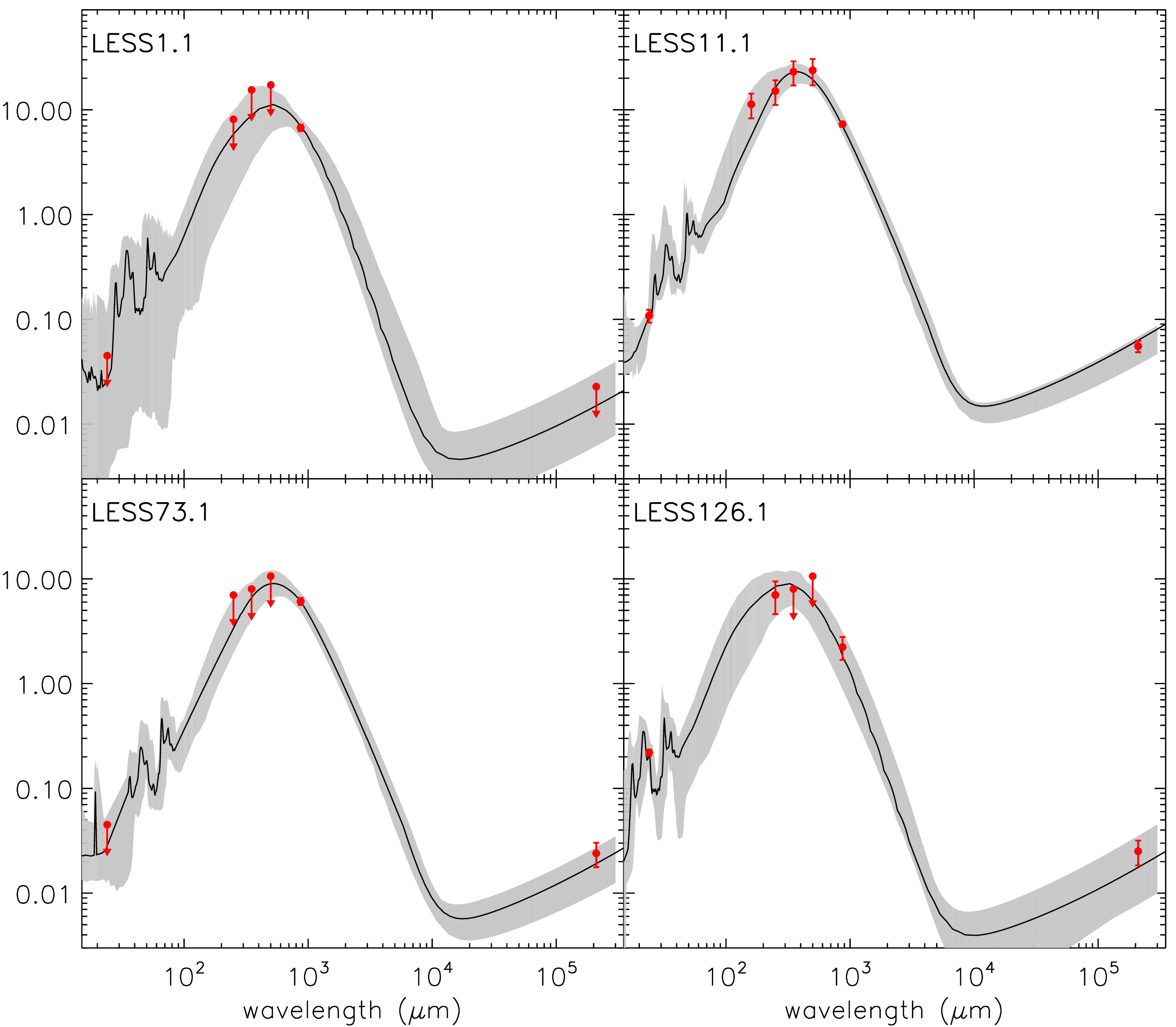,angle=0,width=3.5in}
    \psfig{file=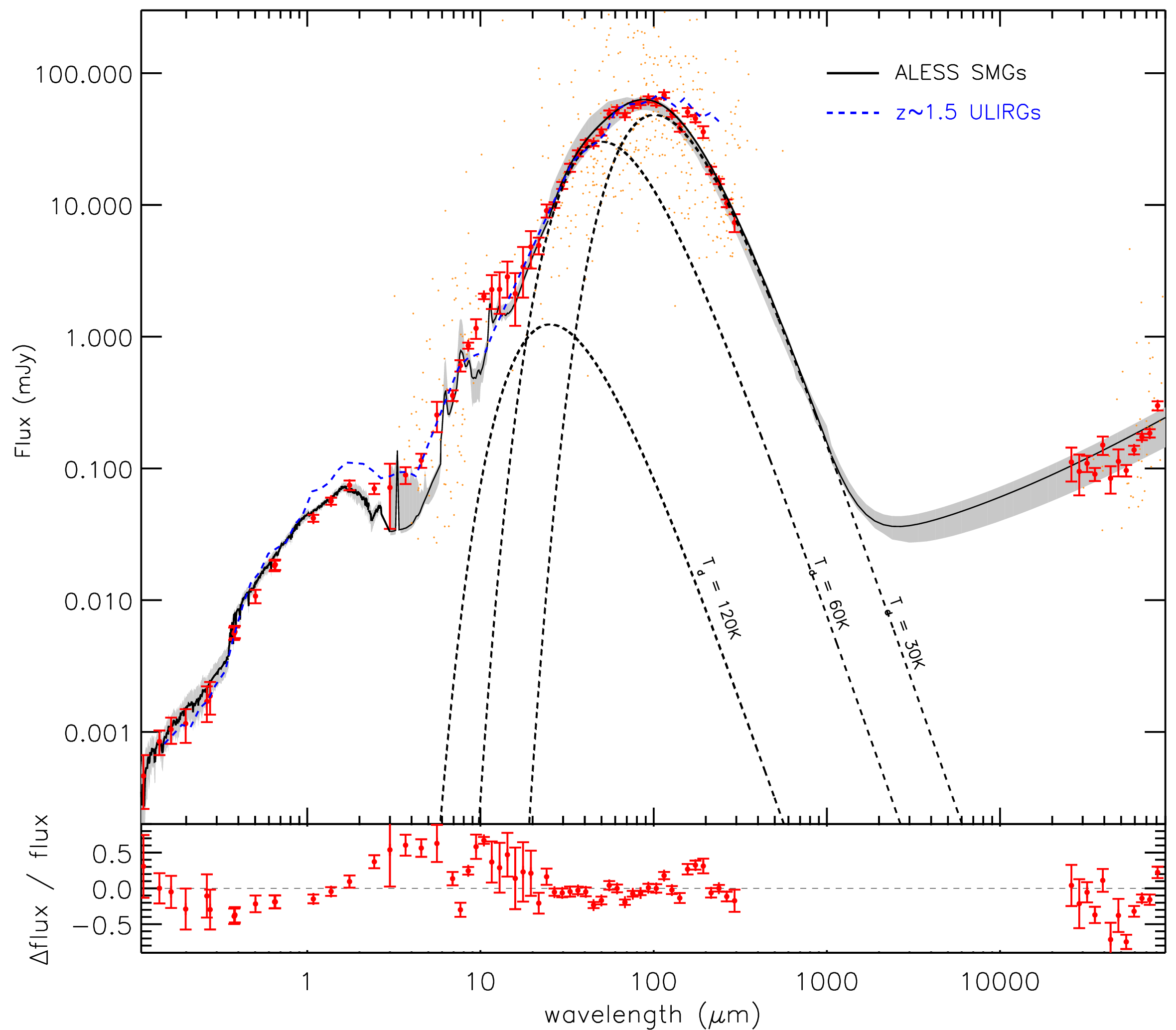,angle=0,width=3.5in}}
\caption{{\it Left:} Example observed mid- to far- infrared SEDs of the
  ALMA SMGs in our sample (one from each quartile in 870$\mu$m LESS
  flux density).  SEDs for all the ALESS SMGs are shown in the
  Appendix.  In each case, the SPIRE photometry has been deblended.
  The solid curve shows the best-fit SED to the 24$\mu$m--1.4\,GHz flux
  densities.  The shaded region shows the range of acceptable solutions
  of these templates given the photometric redshifts (and its
  error). {\it Right:} Rest-frame, composite SED for all ALMA SMGs in
  our sample from UV through to radio wavelengths.  The small points
  show the individual measurements (and includes detections and
  non-detections as limits).  Large points denote the bootstrap median
  in bins of wavelength, with error bars accounting for both
  photometric redshift and luminosity uncertainties.  The solid curve
  shows the best-fit SED, with the 1-$\sigma$ uncertainty indicated by
  the shaded region, and the lower panel shows the residuals between
  the data and the fit.  The dashed curve shows the composite SED
  derived from 816 $z\sim $\,1.5 galaxies with luminosities $L_{\rm
    IR}$\,=\,1--3\,$\times$\,10$^{12}$\,L$_{\odot}$ in the COSMOS field
  from Lee et al.\ (2013).  The black dashed curves show a 3-component
  grey-body dust SED fit to the ALESS SMG composite with cold, warm and
  hot components with $T_{\rm d,c}$\,=\,20--30\,K, $T_{\rm
    d,w}$\,=\,50--60\,K and $T_{\rm d,h}$\,=\,80--120\,K respectively.
  These grey bodies suggest an average cold dust mass of $M_{\rm
    d,c}$\,=\,(4.1\,$\pm$\,0.6)\,$\times$\,10$^{8}$\,M$_{\odot}$ (for a
  dust mass absorption coefficient of $\kappa_{\rm 870\mu
    m}$\,=\,0.15\,m$^2$\,kg$^{-1}$).}
\label{fig:SEDs}
\end{figure*}

\section{Results \& Discussion}
\label{sec:analysis}

\subsection{Average far-infrared colours of SMGs}
\label{sec:stacking}

Before discussing the far-infrared colours of individual ALESS SMGs, we
first investigate the average properties of the galaxies in our sample
by stacking the multi-wavelength photometry in the ``raw''
(non-deblended) maps.  In the PACS and SPIRE maps, we first subtract
the mean flux of 1000 random positions in the map, effectively removing
any systematic contribution from the background or confusion.  We then
stack the far-infrared and radio maps at the positions of the 99 ALESS
SMGs and show these results in Fig.~\ref{fig:HSOstack} and report the
average flux densities in Table~1.  For the entire sample of 99 ALESS
SMGs in the {\sc main} catalog, the composite SED peaks at 350$\mu$m
with flux density ratios of $S_{\rm 250\,\mu m}$\,/\,$S_{\rm 350\,\mu
  m}$\,=\,0.8\,$\pm$\,0.1 and $S_{\rm 350\,\mu m}$\,/\,$S_{\rm 500\,\mu
  m}$\,=\,1.1\,$\pm$\,0.1, as expected for a starburst galaxy with a
dust SED with characteristic dust temperature $\sim $\,30\,K redshifted
to $z\sim $\,2.

To derive the far-infrared properties of this composite dust SED, we
construct a library of local galaxy templates from
\citet{CharyElbaz01,Rieke09} and \citet{Draine07} and also add the SEDs
of the well studied high-redshift starbursts SMM\,J2135$-$0102
($z$\,=\,2.32) and GN\,20 ($z$\,=\,4.05) from \citet{Ivison10eyelash}
and \citet{Carilli11} respectively which can then be fit to the
far-infrared photometry.  This library comprises a total of 185
templates with a range of characteristic dust temperatures (measured
from the wavelength of the peak of the far-infrared SED and assuming
$\lambda_{\rm peak}T_{\rm d}$\,=\,2.897\,$\times$\,10$^{-3}$m.K) from
$T_{\rm d}$\,=\,20--60\,K (this complilation of templates is available
at the same url as the deblended catalogs).

We fit these template SEDs to the 24\,$\mu$m--1.4\,GHz photometry of
the composite SED using a $\chi^2$ minimisation, allowing the
normalisation and redshift of the templates to vary.  In
Fig~\ref{fig:HSOstack} we overlay the best-fit (and also show the range
of models which lie within 1-$\sigma$ of the best fit).

A large fraction of the sub-mm sources identified in single dish
observations lack radio (and\,/\,or 24$\mu$m) counterparts, possibly as
a result of the galaxies either lying at higher redshift or having
colder-than-average dust temperatures.  Until now, we have not known
whether the unidentified SMGs have the same redshift distribution (or
are representative) of the radio-identified subset.  To provide a
simple test of whether the individually radio-detected and
radio-undetected subset show evidence for having different far-infrared
colours (which may indicate a redshift bias if their SEDs are similar),
we stack the radio-detected (46) and radio non-detected (53) subsets
separately and also show these in Fig.~\ref{fig:HSOstack} (see also
Table~1).  Both of these subsets peak at 350$\mu$m, with statistically
indistinguishable 350\,/\,250-$\mu$m and 350\,/\,500-$\mu$m colours.
However, as can be seen from Fig.~\ref{fig:HSOstack} and Table~1, the
70--100$\mu$m flux densities for the radio non-detected stack are
fainter than the radio-detected subset.  For a fixed characteristic
dust temperature, this may be consistent with the radio-faint subset of
the ALESS SMGs lying at somewhat higher redshift.  We will return to a
discussion of this in \S~\ref{sec:FIRprops}.

\subsection{SMGs Redshifts and Rest-Frame Composite SED}
\label{sec:photz}

Recently, Simpson et al.\ (2013) used the 16-band optical--mid-infrared
photometric coverage of ECDFS to derive photometric redshifts for 77 of
the 99 ALESS SMGs.  They fitted SEDs to the 19-band (observed $U$ to
IRAC 8.0$\mu$m) photometry using {\sc hyper-z} with the spectral
templates of \cite{Bruzual03} which use using solar metalicities and
employ four SFHs; a single burst (B), constant star for- mation (C) and
two exponentially decaying SFH’s with timescales of 1 Gyr (E) and 5 Gyr
(Sb).  To calibrate the redshifts, Simpson et al.\ (2013) derive
photometric redshifts of $\sim $\,6000 spectroscopically confirmed
gakaxies in the ECDFS, as well for spectroscopically confirmed ALESS
SMGs (Danielson et al.\ in prep).  Simpson et al.\ (2013) then use the
relation between rest-frame $H$-band magnitude and redshift for ALESS
SMGs below $z$\,=\,2.5 to crudely estimate the redshifts for a further
19 ALESS SMGs which are faint (or undetected) in the
optical\,/\,near-infrared, deriving a statistical redshift distribution
for 96 ALESS SMGs (we note that two of the ALESS SMGs which are faint
or blank in the optical/near-infrared have been confirmed to be
$z$\,=\,4.4 through blind identification of [C{\sc ii}];
\citealt{Swinbank12c}).  The final three ALESS SMGs from the {\sc main}
catalog of \citet{Hodge13} without photometric redshifts lie outside the
MUSYC field and are only covered by IRAC, making it impossible to
derive reliable photometric redshifts.  Simpson et al.\ (2013) show
that the redshift distribution of the 96 ALESS SMGs is centered at
$z$\,=\,2.5\,$\pm$\,0.2 but with a tail out to $z\gsim $\,5, and with a
typical uncertainty for any SMG of $\Delta z$\,/\,(1\,+\,$z_{\rm
  spec}$)\,=\,0.15.  The median absolute $H$-band magnitude of the
ALESS SMGs is $M_{\rm H}$\,=\,$-$24.33\,$\pm$\,0.15 which corresponds
to a stellar mass of $M_{\rm
  star}$\,=\,(8\,$\pm$\,1)\,$\times$\,10$^{10}$\,M$_{\odot}$ (for an
appropriate $L_{\rm H}$\,/\,$M_{\star}$ ratio), consistent with
previous estimates of the stellar masses of SMGs
\citep[e.g.\ ][]{Hainline11}

Simpson et al.\ (2013) used these photometric redshifts to search for
differences in the redshift distribution of radio-bright versus
radio-faint ALESS SMG.  For the radio-detected subset of the
population, they derive a median of $z$\,=\,2.3\,$\pm$\,0.1 whilst for
the radio-undetected subset, they derived $z$\,=\,3.0\,$\pm$\,0.3.  Thus,
it appears that radio-faint SMGs have a redshift distribution which
peaks at slightly higher redshift than the radio-detected SMGs, as
expected given the positive K-correction in the radio wavebands, even
though the 250\,/\,350-$\mu$m and 350\,/\,500-$\mu$m colours are
indistinguishable.  This is discussed in detail in Simpson et
al.\ (2013).

We use the photometric redshifts for the 96 ALESS SMGs to derive a
rest-frame UV--radio composite SED for the whole sample.  For each SMG,
we de-redshift the wavelength and flux density measurements (and
normalise each SMG by far-infrared luminosity) and then calculate a
running median and show this in Fig.~\ref{fig:SEDs}.  To account for
the errors on the SED at each wavelength, we bootstrap resample for
both the photometric redshift and photometric errors.  The best-fit
template (and 1$\sigma$ error distribution) is also overlaid onto the
SED in Fig.~\ref{fig:SEDs} which shows that the best-fit template peaks
at $\lambda_{rest}$\,=\,90\,$\pm$\,5\,$\mu$m.  In this plot, we also
overlay the composite optical--far-infrared SED derived from a sample
of $z\sim $1.5 ULIRGs with luminosities $L_{\rm
  IR}$\,=\,1--3\,$\times$\,10$^{12}$\,L$_{\odot}$ identified in the
COSMOS field from Lee et al.\ (2013).  This composite SED is reasonably
well matched to the ALESS SMG composite, although shows a factor $\sim
$\,1.5--2 excess in the rest-frame near- and mid-infrared compared to
the ALESS stack (which may be due to their sample being dominated by a
24$\mu$m pre-selection).  Nevertheless, in the rest-frame UV/optical
and far-infrared, both SEDs are well matched, peaking between 90--100\,$\mu$m.

This well-sampled rest-frame composite SED can be used to 
measure the average dust masses of the ALESS SMGs.  The dust
mass is related to the far-infrared flux density by $S_{\rm
  \nu}$\,=\,$\kappa_{\nu}$\,$B_{\nu}$($T$)\,$M_{\rm d}$\,$d_{\rm
  L}^2$\,(1\,+\,$z$), where $S_{\nu}$ is the flux density at frequency
$\nu$; $B_{\nu}(T)$ is the Planck function at temperature, T;
$\kappa_{\rm \nu}$ is the dust absorption coefficient; $M_{\rm d}$ is
the total dust mass and $d_{\rm L}$ is the luminosity distance.  Our
sources are not perfect black bodies, but this is accounted for by the
dust mass coefficient, $\kappa_{\nu}$ so that the grey-body is
effectively represented by the product $\kappa_{\nu}\,B_{\nu}(T)$ and
the luminosity of the sources at frequency $\nu$ scales as
$S_{\nu}$\,/\,$B_{\nu}$($T$)\,$\propto$\,$\nu^2$.  The dust mass is
then given by\\ $M_{\rm d}$\,=\,$S_{\rm \nu}$\,d$_{\rm L}^2$ /
($\kappa_{\nu}$\,$B_{\nu}$($T$)\,(1\,+\,$z$)).

To characterise the rest-frame composite ALESS SED in
Fig.~\ref{fig:SEDs}, we fit three dust components; cold: $T_{\rm
  d,c}$\,=\,20--30\,K; warm: $T_{\rm d,w}$\,=\,50--60\,K; and hot
$T_{\rm d,h}$\,=\,80--120\,K.  The dust emissivity, $\beta$ is allowed
to vary between $\beta$\,=\,1.5--2.0 \citep{Magnelli12} (but is the
forced to the same value for each component).  From the best fit, we
derive an average cold dust mass of $M_{\rm
  d,c}$\,=\,(4.1\,$\pm$\,0.6)\,$\times$\,10$^{8}$\,M$_{\odot}$ (for a
dust mass absorption coefficient of $\kappa_{870\mu
  m}$\,=\,0.15\,m$^2$\,kg$^{-1}$;
\citealt{WeingartnerDraine01,Dunne03}) and a ratio of $M_{\rm
  d,c}$\,/\,$M_{\rm d,w}$\,=\,30\,$\pm$\,8 and $M_{\rm
  d,w}$\,/\,$M_{\rm d,h}> $\,1500.  The cold dust mass we derive from
this composite is comparable to that derived for the 
spectroscopically confirmed SMGs,  $M_{\rm
  d}$\,=\,(3.9\,$\pm$\,0.5)\,$\times$\,10$^{8}$\,M$_{\odot}$
\citep{Magnelli12}.  However, this composite SED is derived for all
ALESS SMGs (over all photometric redshifts) and so to limit any
broadening of the dust SED due to selecting different SEDs at different
redshifts, we also limit the composite to the redshift range
$z$\,=\,1.8--2.8.  From this composite, we derive a cold dust mass of
$M_{\rm d,c}$\,=\,(4.1\,$\pm$\,0.6)\,$\times$\,10$^{8}$\,M$_{\odot}$
and a ratio of $M_{\rm d,c}$\,/\,$M_{\rm d,w}$\,=\,43\,$\pm$\,15 and
$M_{\rm d,w}$\,/\,$M_{\rm d,h}> $\,1000 -- comparable to those derived
from the composite SED of all ALESS SMGs.

\subsection{Far-infrared properties of individual SMGs}
\label{sec:FIRprops}

Next, we investigate the properties of individual ALESS SMGs from the
deblended SPIRE maps.  In Fig.~\ref{fig:SEDs} we show example SEDs for
four of the ALESS SMGs in our sample (SEDs for all ALESS SMGs can be
found in Fig.~\ref{fig:allSEDs}).  In cases of non-detections, we show
3$\sigma$ upper limits.  For each ALESS SMG, we fit the SED templates
in our library to the photometry, allowing the redshift to vary
according to the photometric redshift and its error, and also
accounting for the uncertainty in the photometry.  Using the best-fit
dust SEDs, we calculate the infrared luminosity ($L_{\rm IR}$) by
integrating the rest-frame SED between 8--1000$\mu$m (rest-frame).
The derived far-infrared luminosities (integrated between rest-frame
8--1000\,$\mu$m) and characteristic dust temperatures ($T_{\rm d}$) of
the best fit template from these fits are reported in Table~2 along
with their photometric redshifts from Simpson et al. (2013).  To
facilitate a useful comparison with other surveys, we also fit the
far-infrared photometry of each ALESS SMG with a modified black-body
spectrum at the photometric redshift and derive the characteristic dust
temperature from these fits.  These dust temperatures are also reported
in Table~2 and are those used in the analysis below.

\begin{figure*}
  \centerline{\psfig{file=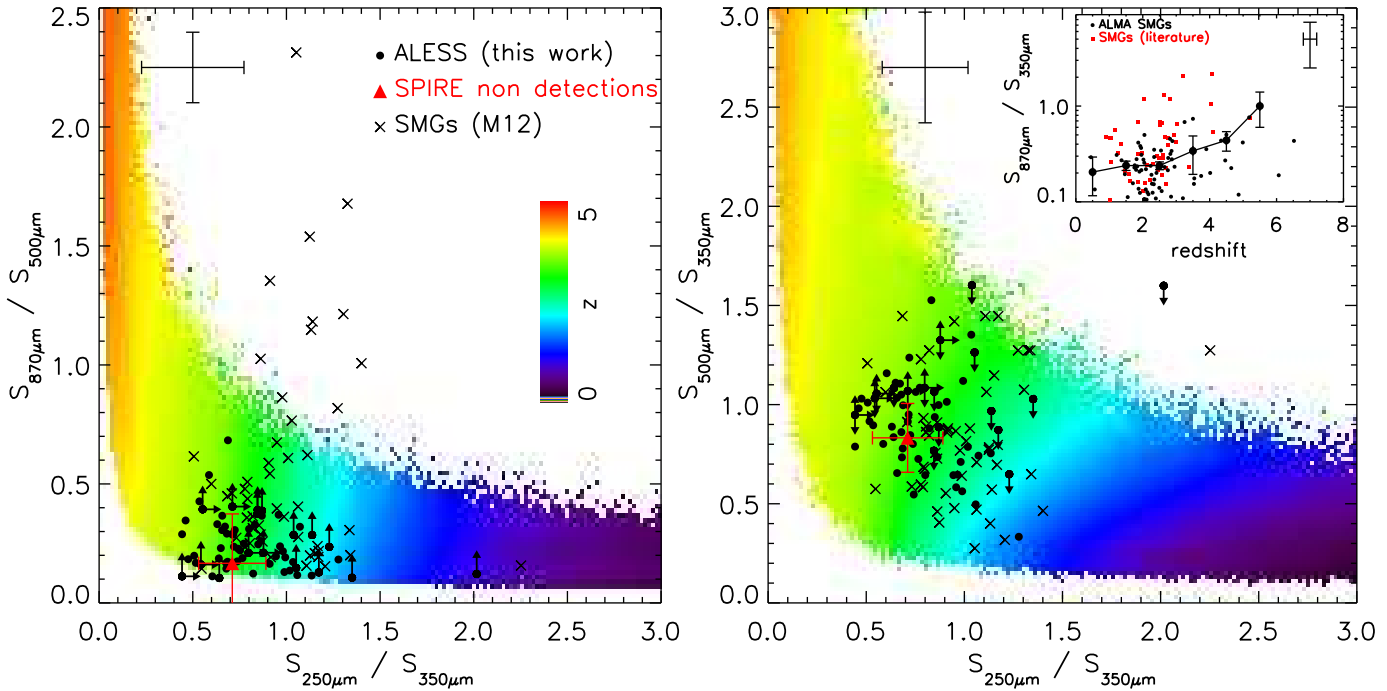,angle=0,width=7in}}
\caption{Far-infrared colours for the ALESS SMGs.  The underlying
  colour scale shows the expected distribution (coded by photometric
  redshift) for a sample of 10$^7$ grey-body curves with a range of
  temperature $T_{\rm d}$\,=\,15\,--\,60\,K, dust emissivity,
  $\beta$\,=\,1.0--2.5 and redshift \emph{z}=\,0\,--\,6.  To account
  for photometry errors in the SPIRE data, these dust SEDs have 10\%
  flux density errors added to the photometry before deriving the
  colours.  In the left panel we show the 250\,/\,350-$\mu$m versus
  870\,/\,500\,-$\mu$m colours of ALESS SMGs and in the right panel we
  show their 250\,/\,350-$\mu$m versus 500\,/\,350\,-$\mu$m colours.
  We only include ALESS SMGs which are detected in at least one SPIRE
  band (65/99 galaxies).  The average colours of the remaining 34 ALESS
  SMGs are shown by the solid triangle using the stacking results in
  Fig.~\ref{fig:HSOstack}.  In both panels, we show a representative error
  bar on the photometry in the top left hand corner.  As evident from
  the figure, the ALESS SMGs have colours consistent with $z\sim
  $\,2--4 dust SEDs.  {\it Inset:} The 870\,/\,350\,-$\mu$m colours as
  a function of photometric redshift for the ALESS SMGs.  In this plot,
  we also include the far-infrared colours of the spectroscopically
  confirmed SMGs from \citet[][M12]{Magnelli12}.  In the top right corner of this inset we show
  a representative error bar on the 870\,/\,350\,$\mu$m colour and
  redshift.  The solid line shows the median (and error) in $\Delta
  z$\,=\,1 bins.  This shows that there is a scatter of approximately
  $\Delta z\gsim $\,1 for a fixed 870\,/\,350\,$\mu$m colour.}
\label{fig:FIRcolours}
\end{figure*}

Following \citet{Ivison12}, in Fig.~\ref{fig:FIRcolours} we show the
far-infrared (250\,/\,350\,$\mu$m versus 500\,/\,350\,$\mu$m and
870\,/\,500\,$\mu$m) colours of the ALESS SMGs (we only plot ALESS SMGs
which are detected in at least two bands).  For a comparison sample, we
also include the far-infrared colours of SMGs
with 250, 350 and 500$\mu$m flux densities measured from
\citet{Magnelli12}.  This colour-colour diagnostic is designed to
crudely assess the redshift and characteristic dust temperature
($T_{\rm d}$) of galaxies detected by \emph{Herschel}, probing their
colours across the rest-frame $\sim $\,100$\mu$m SED peak.

To assess whether these colours are consistent with those expected for
dusty high-redshift galaxies, we also show as a colour scale the
expected far-infrared colours derived from 10$^6$ grey body curves with
a range of redshifts from $z$\,=\,0--6, characteristic dust
temperatures of $T_{\rm d}$\,=\,15--60\,K and dust emissivity
$\beta$\,=\,1.0--2.5 (we include scatter in these photometry which
match the typical photometric errors in our analysis).  The location of
the ALESS SMGs in Fig.~\ref{fig:FIRcolours} demonstrates that their
dust SEDs are consistent with a population of galaxies at $z\sim
$\,2--4, and we note that there are 12, 32 and 12 ALESS SMGs whose dust
SEDs peak closest to 250, 350 and 500\,$\mu$m respectively (these are
for those galaxies which are individually detected in at least two
SPIRE bands).  However, due to the dust-temperature--redshift
degeneracy, there is significant scatter between the far-infrared
colours and photometric redshift.  Indeed, as also shown in
Fig.~\ref{fig:FIRcolours} the relation between 870\,/\,350\,-$\mu$m
colour as a function of photometric redshift for the ALESS SMGs (and
also including the far-infrared colours of the spectroscopically
confirmed SMGs from \citealt{Magnelli12}), there is approximately
$\Delta z\gsim $\,1 of scatter for a fixed 870\,/\,350\,-$\mu$m colour.

\begin{figure}
  \centerline{
    \psfig{file=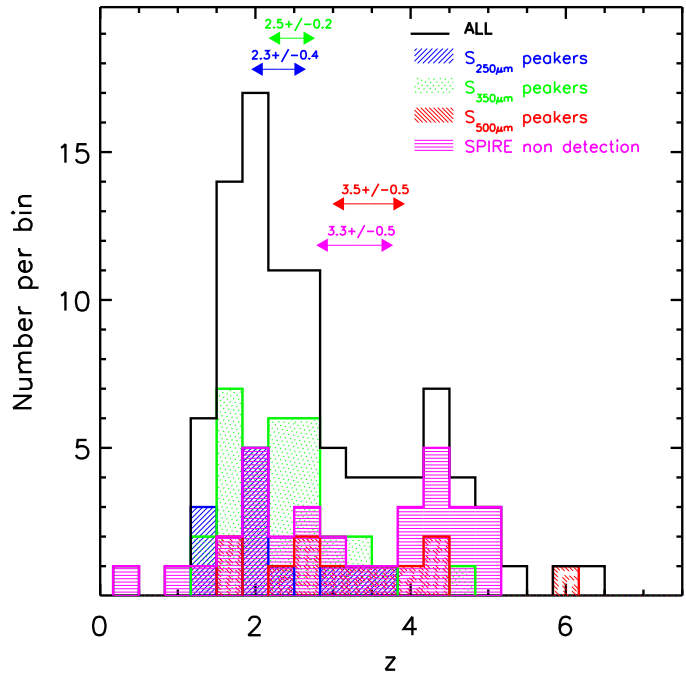,angle=90,width=3.5in}}
\caption{Photometric redshift distribution for the ALESS SMGs.  The
  full distribution peaks at $z$\,=\,2.5\,$\pm$\,0.2, with a tail to
  $z> $\,5 (Simpson et al.\ 2013).  In this plot, we also split the
  SMGs in to those whose observed dust SEDs peak closest to 250, 350
  and 500$\mu$m, deriving median redshifts of $z$\,=\,2.3\,$\pm$\,0.3,
  2.5\,$\pm$\,0.2 and 3.5\,$\pm$\,0.5 respectively (for galaxies
  individually detected at 350$\mu$m).  We also plot the photometric
  redshift distribution for the ALESS SMGs which do not have a
  $>$3\,$\sigma$ counterpart at 250, 350 or 500$\mu$m, deriving a
  median redshift of $z$\,=\,3.3\,$\pm$\,0.5.  This plot demonstrates
  that those galaxies with SEDs that peak at longer far-infrared
  wavelength, or are undetected by SPIRE, tend to lie at higher
  redshifts, although there is considerable overlap between the
  samples.}
\label{fig:Nz}
\end{figure}

In
Fig.~\ref{fig:Nz} we show the photometric redshift distribution for
ALESS SMGs, split by their far-infrared colours.  Crudely, for a fixed
temperature, the dust SEDs for the SMGs which peak at shorter
wavelengths should lie at the lower redshifts, whilst those which peak
at the longer wavelengths should lie at the highest redshifts.  As
Fig.~\ref{fig:Nz} shows, this is broadly consistent with our data; the
dust SEDs of the ALESS SMGs which peak closest to 250, 350 and
500$\mu$m peak at $z$\,=\,2.3\,$\pm$\,0.2, 2.5\,$\pm$\,0.3 and
3.5\,$\pm$\,0.5 respectively.  Formally, a two-sided KS test suggests a
63\% chance the the 250-$\mu$m and 350-$\mu$m peakers are drawn from
the same distribution, but only a 2.3\% [1.8\%] chance that the
350-$\mu$m and 500-$\mu$m [250-$\mu$m and 500-$\mu$m] peakers are drawn
from the same population.

Finally, we note that there are 34 (out of 99) ALESS SMGs which do not
have a $>$3\,$\sigma$ counterpart at 250, 350 or 500$\mu$m.  Of these
34 galaxies, 30 are also radio unidentified, and it is interesting to
note that the median photometric redshift for these SPIRE and radio
non-detections is higher than the full ALESS SMG sample, with
$z$\,=\,3.3\,$\pm$\,0.5 (c.f.\ $z$\,=\,2.5\,$\pm$\,0.2; Simpson et
al.\ 2013; Fig.~\ref{fig:Nz}).  However, stacking the SPIRE maps of
these ``SPIRE undetected'' ALESS SMGs (Fig.~\ref{fig:SEDs}) yields
far-infrared colours which peak at 350$\mu$m with 250, 350 and
500$\mu$m flux densities of $S_{\rm 250\mu
  m}$\,=\,9.0\,$\pm$\,0.4\,mJy, $S_{\rm 350\mu
  m}$\,=\,9.5\,$\pm$\,0.5\,mJy and $S_{\rm 500\mu
  m}$\,=\,6.5\,$\pm$\,1.2\,mJy (Fig.~\ref{fig:FIRcolours}).  Moreover,
 the median 870$\mu$m flux density of this ``SPIRE undetected'' subset
is $S_{\rm 870\mu m}$\,=\,2.4\,$\pm$\,0.4\,mJy, (c.f.\ $S_{\rm 870\mu
  m}$\,=\,3.4\,$\pm$\,0.3\,mJy for the full ALESS SMG sample).  Thus,
these ALESS SMGs which are undetected in the SPIRE maps appear to
represent a combination of the slightly fainter and higher redshift
subset of the ALESS SMGs, but with comparable dust temperatures.  Of
course, this result may partially due to our method of deblending in
which we require a galaxy be detected at 250-$\mu$m to be included in
the 350-$\mu$m positional prior catalog (and a 350-$\mu$m to be
included in the 500-$\mu$m positional prior catalog) which may bias
against galaxies that are faint at 250-$\mu$m and peak at longer
wavelengths.  However, we note that only two ALESS SMGs, ALESS\,80.1
and ALESS\,80.2 ($z_{\rm phot}$\,=\,1.4 and 2.0 respectively) have
significant flux ($\gsim $\,8\,mJy at 350-$\mu$m and 500-$\mu$m) in the
residual map, and therefore it does not appear that we have missed a
significant fraction of ALESS SMGs in the SPIRE deblending which would
bias this result.

For each ALESS SMG, we also calculate the cold dust mass using the
rest-frame 870$\mu$m luminosity from the best-fit template and give
these in Table~2.  The median dust mass for all of the SMGs in our
sample is $M_{\rm
  d}$\,=\,(3.6\,$\pm$\,0.3)\,$\times$\,10$^{8}$\,M$_{\odot}$, which is
comparable to the average dust mass derived by \citet{Magnelli12} for a
spectroscopic sample of SMGs, $M_{\rm
  d}$\,=\,(3.9\,$\pm$\,0.5)\,$\times$\,10$^{8}$\,M$_{\odot}$.

\section{Discussion}
\label{sec:discussion}

\subsection{Far-Infrared Luminosities \& Star Formation Rates}
\label{sec:SFR}

The median infrared luminosity for the ALESS SMGs is $L_{\rm
  IR}$\,=\,(3.0\,$\pm$\,0.3)\,$\times$\,10$^{12}$\,$L_{\odot}$
(corresponding to a star formation rate of
SFR\,=\,310\,$\pm$\,30\,$M_{\odot}$\,yr$^{-1}$ for a Chabrier IMF) with a range of $L_{\rm
  IR}$\,=\,0.2--10\,$\times$\,10$^{12}$\,$L_{\odot}$
(SFR\,=\,20--1030\,M$_{\odot}$\,yr$^{-1}$).  This is a factor
$\sim$\,1.8$\times$ lower than that derived for 78 spectroscopically
confirmed SMGs from \citet{Chapman05a} (also derived using SPIRE
photometry to constrain the dust SEDs;
SFR\,=\,500\,$\pm$\,66\,M$_{\odot}$\,yr$^{-1}$ and for the same stellar
IMF; \citealt{Magnelli12}).  However, this is mainly driven by the high
multiplicity of SMGs due to unresolved companions in the far-infrared
photometry in the single dish survey.  For example, \citet{Hodge13}
show that at least 35\% (possibly up to 50\%) of the LABOCA sub-mm
sources are resolved by ALMA into multiple SMGs (see also
\citealt{Karim13}).  If we instead limit the sample to ALESS SMGs whose
flux is brighter than $S_{\rm 870}\geq $\,4.2\,mJy, then we derive an
average star formation rate of
SFR\,=\,530\,$\pm$\,60\,M$_{\odot}$\,yr$^{-1}$.

\begin{figure*}
  \centerline{\psfig{file=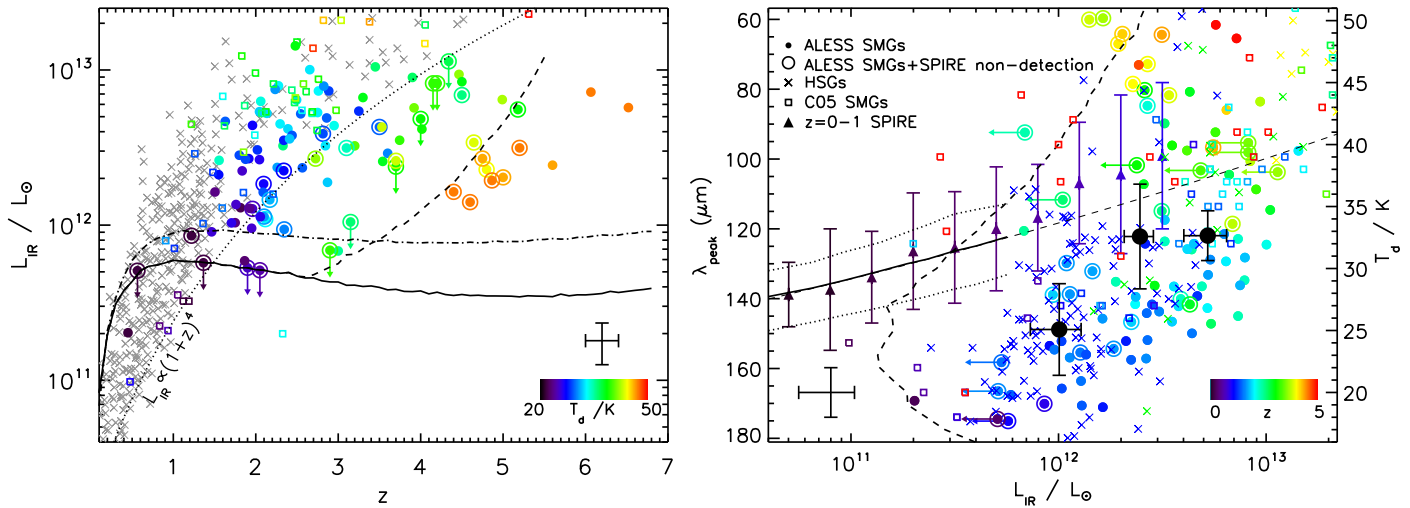,angle=0,width=7in}}
\caption{ {\it Left:} The infrared luminosity as a function of
  photometric redshift for the ALESS SMGs, colour coded by
  characteristic dust temperature.  The dotted line denotes luminosity
  evolution according to $L_{\rm IR}\propto $\,(1\,$+$\,$z$)$^4$.  We
  also include several selection functions: {\it (i)} selection of an
  $S_{\rm 870\mu m}>$\,2\,mJy SMG for a dust SED with characteristic
  dust temperature of $T_{\rm d}$\,=\,32\,K (solid line); {\it (ii)}
  selection of an $S_{\rm 870\mu m}>$\,2\,mJy SMG for a dust SED with
  characteristic dust temperature that evolves with luminosity and
  redshift (dot-dashed line); and {\it (iii)} $S_{\rm 870\mu
    m}>$\,2\,mJy SMGs which are likely to be detected in at least two
  SPIRE bands (dashed line).  {\it Right:} The relation between the
  characteristic dust temperature (which closely corresponds to the
  approximate peak wavelength of the dust SED, $\lambda_{\rm peak}$)
  versus infrared luminosity ($L_{\rm IR}$) for the ALESS SMGs, colour
  coded by photometric redshifts.  To facilitate a useful comparison
  with other surveys, the characteristic dust temperatures shown here
  are derived using the grey-body fits to the far-infrared photometry
  at the photometric redshift of each SMG.  The dashed line denotes the
  approximate selection limits for an 870$\mu$m selected sample with
  $S_{\rm 870\mu m}>$\,2\,mJy which are then detected in at least two
  SPIRE bands.  We also plot the (local) 60-$\mu$m {\it IRAS} $L_{\rm
    IR}$--$T_{\rm d}$ correlation from \citet{Chapman03b} (solid line,
  with 1-$\sigma$ dispersion shown by the dotted line; see also
  \citealt{Chapin09}) and also the $z$\,=\,0--1 SPIRE-selected LIRGs
  and ULIRGs \citep{Symeonidis13}.  In both panels, we also include
  recent measurements for SMGs from \citet{Magnelli12} which are partly
  based on the \citet{Chapman05a} (C05) survey and \emph{Herschel}
  Star-forming Galaxies (HSGs) from \citep{Casey12}.  In both panels,
  we also show a representative error bar for our ALESS measurements.
  The large, solid points in the right hand panel show the median
  temperature (and bootstrap error) of ALESS SMGs in bins of
  far-infrared luminosity, showing that the high redshift SMGs have
  cooler temperatures ($\Delta T$\,=\,3--5\,K, or equivalently, their
  dust SEDs peak at longer wavelengths) at fixed luminosity than
  comparably luminous galaxies at $z$\,=\,0, which is likely due to the
  more extended star formation in these systems.}
\label{fig:Td_FIR_z}
\end{figure*}

In Fig.~\ref{fig:Td_FIR_z} we plot the correlation between redshift and
infrared luminosity for our sample.  For comparison we overlay the SMGs
and 24-$\mu$m selected \emph{Herschel} Star-forming Galaxies (HSGs)
from \citealt{Chapman05a} and \citealt{Casey12} respectively.
To highlight the selection functions on this plot, we calculate the
870$\mu$m flux density for a dust SED with characteristic dust
temperature of $T_{\rm d}$\,=\,32\,K as a function of luminosity and
redshift.  In Fig.~\ref{fig:Td_FIR_z} the solid line denotes the
selection limit for this dust SED with $S_{870\mu m}> $\,2\,mJy (the
approximate 870$\mu$m flux density limit of the ALESS SMG sample).
However, given the apparent evolution in the luminosity--temperature
plane (high-redshift ULIRGs tend to be systematically cooler than local
ULIRGs at a fixed luminosity;
\citealt{Chapman05a,Kovacs06,Symeonidis13}), we also construct a set of
dust SEDs across the redshift range $z$\,=\,0--8 with a range of
far-infrared luminosities from L$_{\rm
  IR}$\,=\,10$^{10-14}$\,L$_{\odot}$ but require that the dust SEDs of
higher redshift galaxies have cooler characteristic dust temperatures
by $\Delta T_{\rm d}$\,/\,$\Delta z$\,=\,1.5\,K at a fixed luminosity.
We then measure the 870$\mu$m flux densities of each dust SED and again
only record those whose 870$\mu$m emission is above $S_{\rm 870\mu
  m}$\,=\,2.0\,mJy.  We also use the same model to estimate the
selection boundary for SMGs to be detected in at least two SPIRE bands
above our limits (see \S~\ref{sec:deblending}); we show this on
Fig.~\ref{fig:Td_FIR_z}.  This final selection function highlights a
steeply rising selection boundary above $z\sim $\,3.5 beyond which
galaxies are unlikely to be detected in at least two SPIRE bands.
Nevertheless, this plot demonstrates that the ALESS SMGs apparently
follow the same luminosity--redshift scaling ($L_{\rm IR}\propto
$(1\,+\,$z$)$^4$) as local samples, although the luminosities appear to
significantly flatten on this relation above $z\sim $\,2.5.

\begin{figure*}
  \centerline{
    \psfig{file=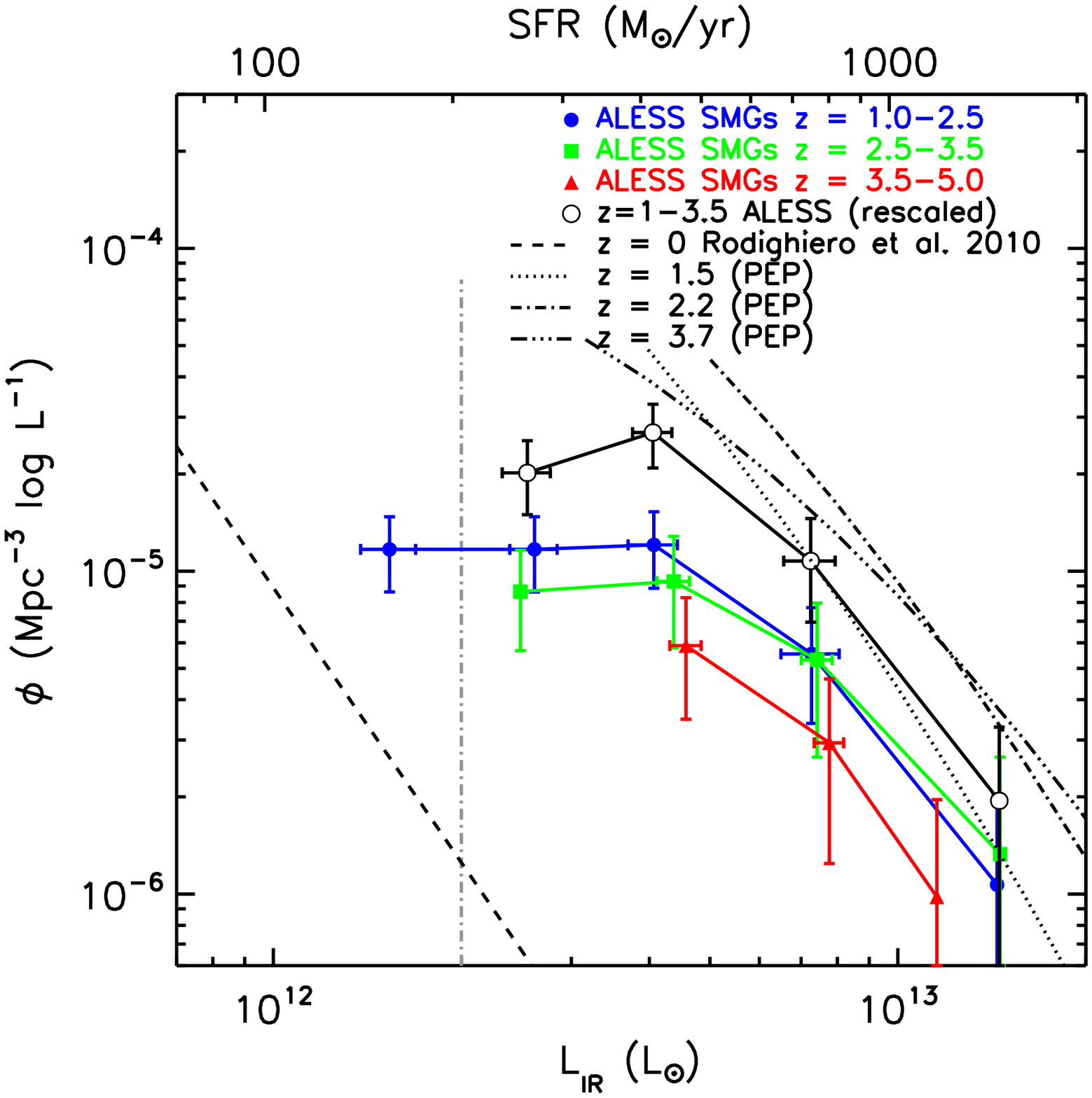,angle=0,width=3.5in}
    \psfig{file=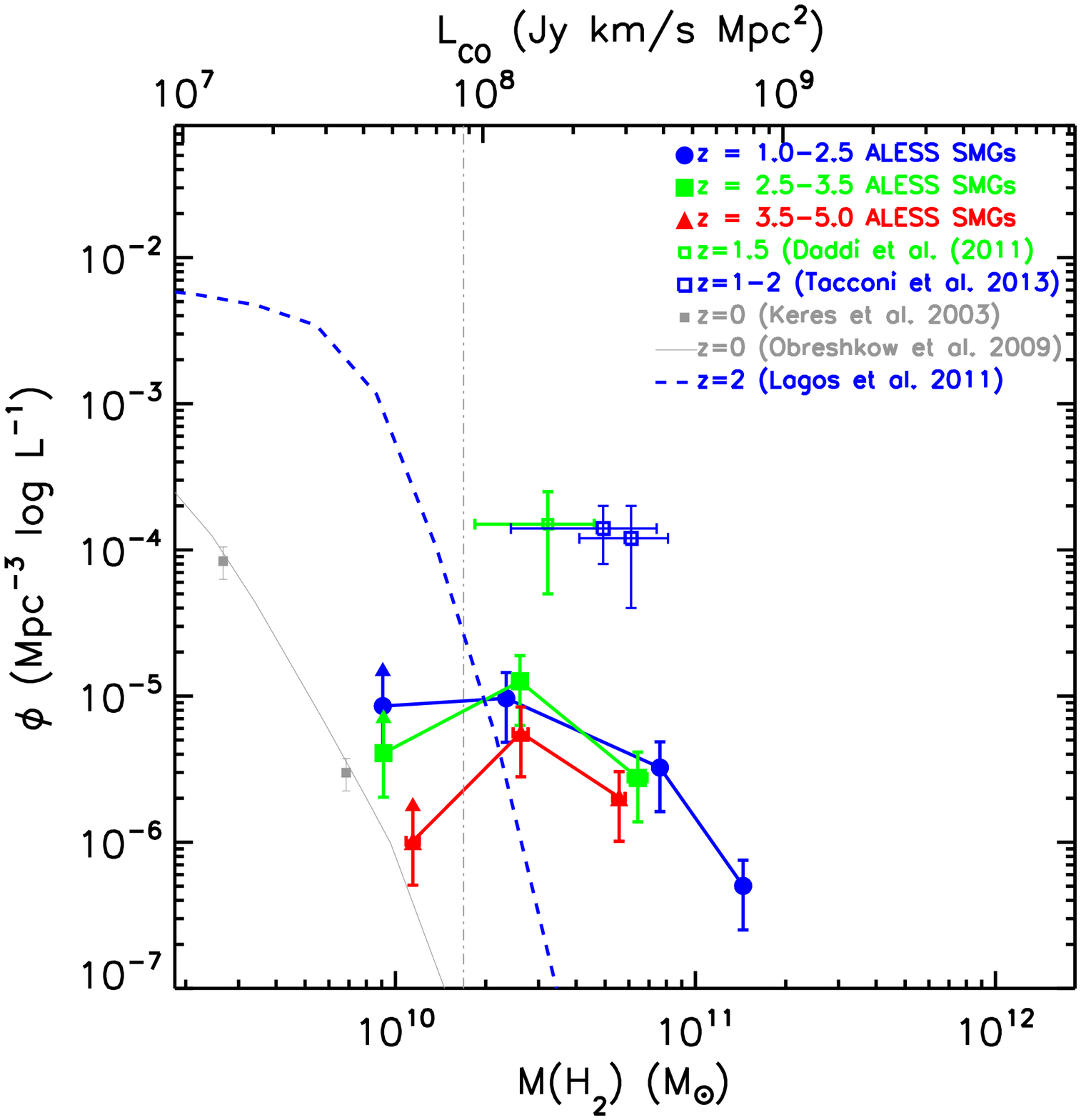,angle=0,width=3.5in}}
\caption{{\it Left:} The far-infrared luminosity function for ALESS
  SMGs split into three bins of photometric redshift.  In this plot,
  the error bars are derived by bootstrap resampling for uncertainties
  in the luminosities and photometric redshifts.  Since the ECDFS has
  been shown to be under-dense in bright sub-mm sources
  \citep{Weiss09}, to match the 870$\mu$m number counts in other
  extra-galactic survey fields, we also re-scaled the $z$\,=\,1.5--3.5
  luminosity function by a factor 2$\times$ (black solid line).  The
  vertical dot-dashed line shows the approximate ALESS completeness
  limit.  We also show the $z$\,=\,0 luminosity function from
  \citet{Rodighiero10} as well as the $z\sim $\,1.5, 2.2 and 3,7
  infrared luminosity function of 100-$\mu$m and 160-$\mu$m selected
  galaxies from the PEP survey from \citet{Gruppioni13} which are well
  matched to the (rescaled) ALESS SMGs.  {\it Right:} The H$_{2}$ mass
  function for SMGs compared to local data and theoretical models.  In
  this plot, we have adopted a gas-to-dust ratio of appropriate for
  each galaxy given its star formation rate and stellar mass, and
  applied the correction factor for the under-density of bright sub-mm
  sources in ECDFS.  We compare the results for the ALESS SMGs with
  $z$\,=\,0 which shows that at a fixed gas mass (or equivalently,
  CO(1-0) line luminosity) there are $\sim$100$\times$ more galaxies at
  $z$\,=\,2 than $z$\,=\,0.  We also include on the plot the space
  density of ``main-sequence'' starburst galaxies (BzKs and BX/BMs)
  from \citet{Daddi10} and \citet{Tacconi12}.  The dashed line shows
  the predicted $z$\,=\,2 gas mass function from \citet{Lagos11}.}
\label{fig:COLF}
\end{figure*}

In Fig.~\ref{fig:Td_FIR_z} we also show the relation between the
infrared luminosity and characteristic dust temperature for ALESS SMGs.
In this plot, we use the characteristic dust temperature from the
grey-body fits in order that a fair comparison can be made with other
surveys.  However, since the characteristic dust temperature is closely
related to the wavelength of the peak of the dust SED, we label the
axis with both dust temperature (T$_{\rm d}$) and corresponding peak
wavelength ($\lambda_{\rm peak}$).  In the plot, we also include
measurements for SMGs from \citet{Magnelli12} as well as the
\emph{Herschel} Star-forming Galaxies (HSGs) from \citet{Casey12}
(which have a median redshift of $z\sim $\,0.7), and the $z$\,=\,0--1
SPIRE-selected LIRGs and ULIRGs from \citet{Symeonidis13} which appear
to closely follow the $z\lsim $\,0.1 $L_{\rm IR}$--$T_{\rm d}$
correlated observed in the $>$\,1.2\,Jy {\it IRAS} 60-$\mu$m sample
(\citealt{Chapman03c}, see also \citealt{Chapin09}).  In this plot, we
show a selection boundary (dashed line) that denotes the
luminosity--temperature space required for an ALESS SMG with $S_{\rm
  870\mu m}> $\,2\,mJy to be detected in at least two SPIRE bands
(which as recently reiterated by \cite{Symeonidis13}, drives the
apparent correlation between $T_{\rm d}$ and redshift).

To test whether the ALESS SMGs have similar temperatures to local
samples (at fixed luminosity), we divide the ALESS SMGs into three
roughly equal number bins of $L_{\rm IR}$ and derive the average
characteristic temperature of $T_{\rm d}$\,=\,25\,$\pm$\,4,
33\,$\pm$\,2 and 33\,$\pm$\,1\,K for $L_{\rm
  IR}$\,=\,(1.0\,$\pm$\,0.1), (2.4\,$\pm$\,1.3) and
(5.3\,$\pm$\,0.4)\,$\times$\,10$^{12}$\,L$_{\odot}$ respectively.  As
Fig.~\ref{fig:Td_FIR_z} shows, for a fixed luminosity, ALESS SMGs have
cooler dust temperatures ($\Delta T_{\rm d}$\,=\,3--5\,K) compared to
that implied from the $L_{\rm IR}$--$T_{\rm d}$ relation from local
60-$\mu$m {\it IRAS} galaxies (equivalently, the dust SEDs of the ALESS
SMGs peak at 10--15$\mu$m shorter wavelengths the local IRAS galaxies
of comparable luminosity).  This offset in $T_{\rm d}$ (or
$\lambda_{\rm peak}$) at fixed $L_{\rm IR}$ for high-redshift ULIRGs
has also been noted by \citet{Symeonidis13} and may be attributed to
the more extended gas and dust distributions and/or higher dust masses
than local galaxies of similar luminosities.

\begin{figure*}
  \centerline{\psfig{file=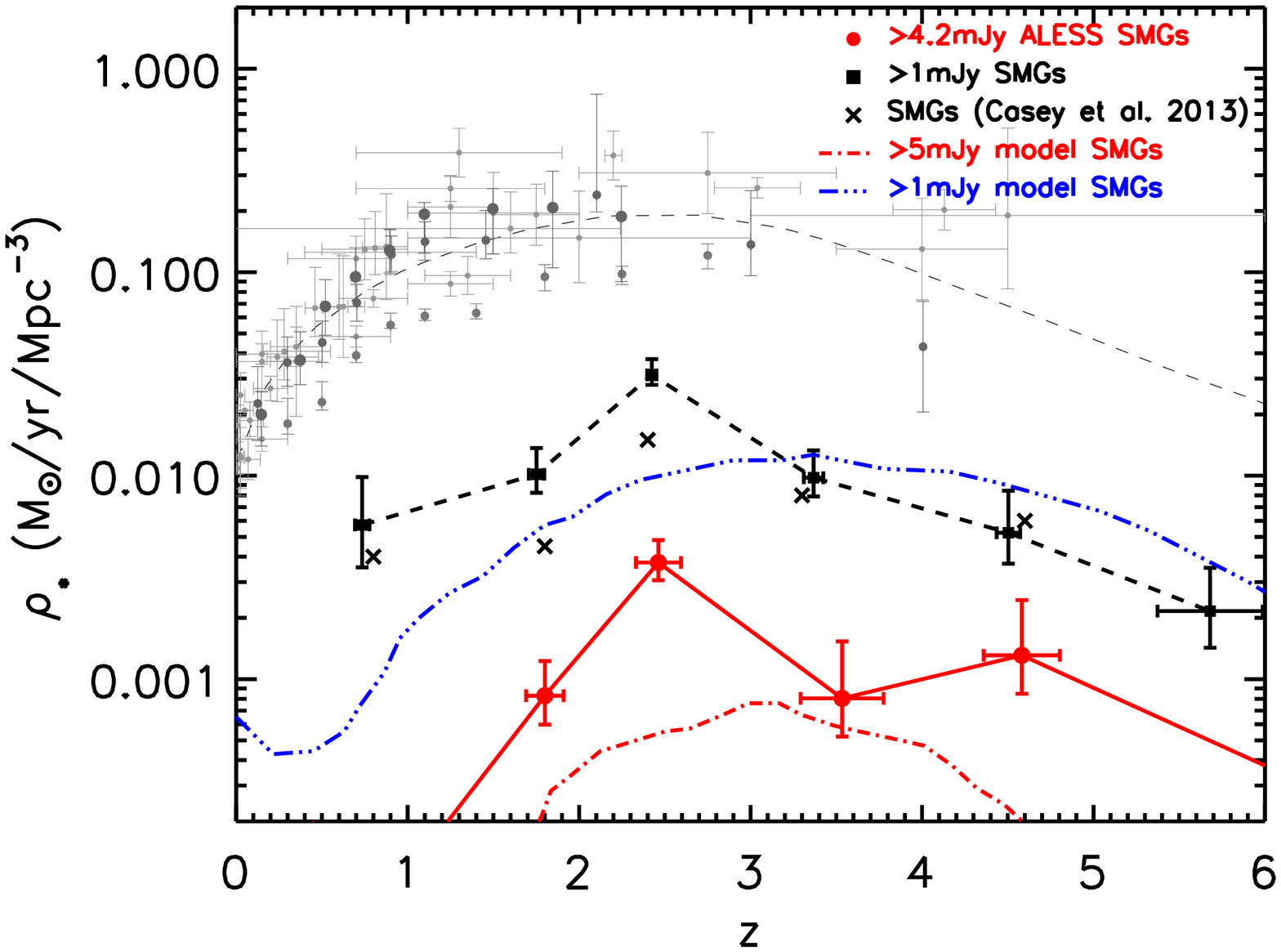,angle=0,width=3.5in}
    \psfig{file=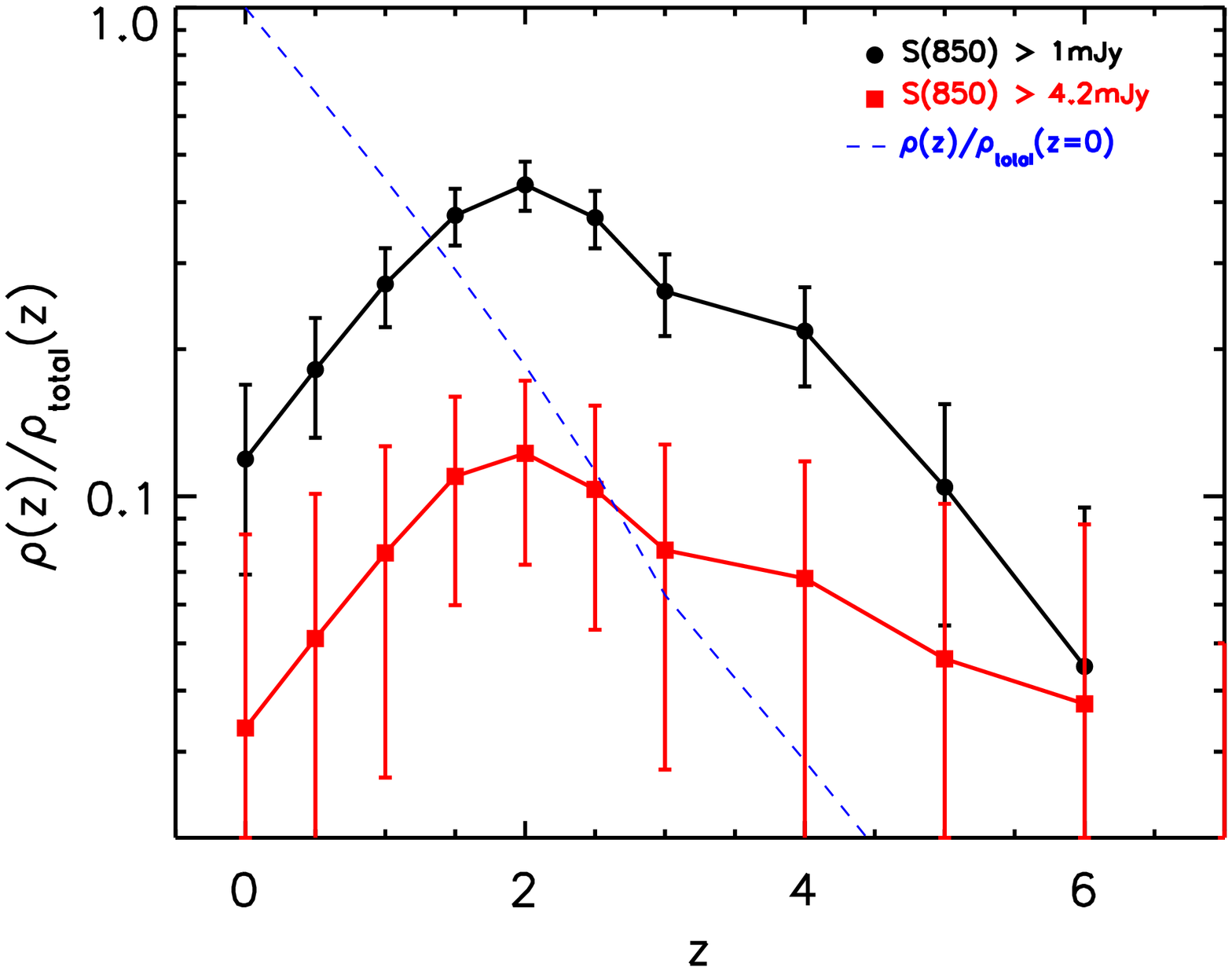,angle=90,width=3.5in}}
\caption{{\it Left:} The contribution of SMGs to the co-moving cosmic
  star formation density as a function of redshift.  Since the original
  LESS survey had flux density limits of $S_{\rm 870\mu
    m}$\,=\,4.2\,mJy, we only include ALESS SMGs brighter this limit,
  but then extrapolate to all SMGs brighter then $S_{\rm 870\mu
    m}$\,=\,1\,mJy using the 850$\mu$m counts \citep{Coppin06}.  This
  plot shows that the SMG activity peaks at $z\sim $\,2 -- similar to
  that found by previous studies of star-forming galaxies and the peak
  activity of QSOs \citep{Hopkins07}.  The contribution from the bright
  SMGs to the total SFRD also peaks at $z\sim $\,2 where they are
  responsible for $\sim $\,1--2\% of the \citet{Hopkins06} SFRD,
  although extrapolating to the faintest SMGs, $\sim $\,1\,mJy suggests
  SMGs contribute up to 20\% of the total SFRD at this epoch. {\it
    Right:} Fraction of stellar mass in SMGs ($S_{\rm 870\mu m}>
  $\,4.2\,mJy) compared to the total stellar mass density as a function
  of redshift.  The global stellar mass densities as a function of
  redshift are taken from \citet{Marchesini10}. This figure shows that
  at $z\gsim $\,2--3 the bright SMGs ($S_{\rm 870\mu m}> $\,4.2\,mJy)
  contribute $\sim $\,15\% of the total stellar mass budget at that
  epoch, and make up 3--4\% of the present day stellar mass density.
  Integrating to fainter luminosities, SMGs with 870$\mu$m fluxes
  brighter than $S_{\rm 870\mu m}> $\,1\,mJy are predicted to
  contribute $\sim$\,30--40\% of the stellar mass density at $z\sim
  $\,2.}
\label{fig:SFRd}
\end{figure*}

\subsection{Far-Infrared Luminosity Functions}

Since our ALMA survey was carried out on a complete sample of
870$\mu$m-selected sources lying in a single field, we can use the
volume probed by our observations to derive the far-infrared luminosity
function of bright SMGs.  To search for an evolution with redshift, we
split the ALESS SMGs in to three bins of photometric redshift
($z$\,=\,1.0--2.5; $z$\,=\,2.5--3.5 and $z$\,=\,3.5--5.0).  To account
for the flux limited nature of the LESS survey, we calculate the ALESS
SMG luminosity function within an accessible volume using
$\phi(L)\Delta L=\Sigma(1\,/\,V_i)$ where $\phi(L)\Delta L$ is the
number density of sources with luminosities between $L$ and
$L$\,+\,$\Delta L$ and $V_i$ is the co-moving volume within which the
$i^{th}$ galaxy can be detected in each luminosity bin.  Error-bars are
calculated by bootstrapping accounting for the uncertainties in the
photometric redshifts, luminosity and binning errors.  In
Fig.~\ref{fig:COLF} we plot the bolometric luminosity function in these
three redshift bins and compare the data to a low-redshift 24-$\mu$m
selected ($z <$\,0.3) sample from \citet{Rodighiero10}, and the $z\sim
$\,1.5, 2.2 and 3.7 infrared luminosity functions of 100 and 160-$\mu$m
selected galaxies from the PEP survey \citep{Gruppioni13}.  As
Fig.~\ref{fig:COLF} shows, at fixed luminosity, SMGs have a space
density at least a factor 100\,$\times$ that of 24-$\mu$m selected
galaxies at $z< $\,0.3 \citep{Rodighiero10}.

Comparing the ALESS SMG space densities between redshifts, we also see
that between $z$\,=\,1.5--2.5 and $z$\,=\,2.5--3.5, the ALESS SMG
luminosity functions significantly overlap, although at a fixed
luminosity, the higher redshift ALESS SMGs tend to have a lower space
density (e.g.\ at a luminosity of $\sim
$\,5\,$\times$\,10$^{12}$\,L$_{\odot}$, the space density of
$z$\,=\,3.5--5.0 SMGs is $\sim$\,60\% lower than at $z\sim $\,2),
implying that the volume density peaks at $z\sim $\,2 and declines at
higher redshift.

As noted by \citet{Weiss09} (see also \citealt{Wardlow11}), the ECDFS
appears to be under-dense in $z\sim $\,2 SMGs by a factor $\sim
$\,2$\times$ compared to other sub-millimetre surveys at flux densities
$S_{\rm 870}\gsim $\,3\,mJy.  To compare to other surveys, in
Fig.~\ref{fig:COLF} we therefore construct the luminosity function for
ALESS SMGs over the redshift range $z$\,=\,1--3.5 (to match the
redshift distribution of the \citealt{Chapman05a} sample) and then
rescale by increasing the number density by a factor $\sim
$\,2\,$\times $ to match the 850$\mu$m number counts from those in the
SHADES \citep{Coppin06}.  This rescaled $z$\,=\,1--3.5 ALESS SMG
luminosity function is well matched to the $z\sim $\,2.2 luminosity
functions of 100 and 160-$\mu$m selected galaxies from the PEP survey
\citep{Gruppioni13}.

\subsection{The contribution of SMGs to the co-moving star formation rate and stellar mass densities}

We use the star formation rates for the galaxies in our sample to
measure the contribution of SMGs to the total star formation rate
density as a function of redshift.  When calculating the star formation
rate density, we include all of the SMGs from the ALESS {\sc main}
catalog with $S_{870}>$\,4.2\,mJy (the flux limit of the original LESS
survey), and account for the factor 2\,$\times$ under-density of SMGs
in the ECDFS.  As Fig.~\ref{fig:SFRd} shows, over the redshift range
$z$\,=\,1--5, bright (S$_{\rm 870} >$\,4.2\,mJy) account for $\sim
$\,1--2\% of the total star formation density
\citep{Hopkins06,Karim11}.  
However, we also need to account for the large fraction of the
sub-millimeter galaxy population below our bright $\sim$\,4.2\,mJy flux
limit, by integrating down to 1\,mJy.  This flux represents the point
at which the dust-obscured and unobscured star formation rates in
galaxies are comparable and corresponds to an infrared luminosity of
$L_{\rm IR}$\,=\,0.8\,$\times$\,10$^{12}$\,L$_{\odot}$ (SFR\,$\sim
$\,80\,M$_{\odot}$\,yr$^{-1}$).  We assume that the fainter SMGs
($S_{\rm 870\mu m}$\,=\,1--4.2\,mJy) have the same underlying redshift
distribution and luminosity evolution as the bright SMGs ($S_{\rm
  870\mu m} >$\,4.2\,mJy) and find that the number density of SMGs with
870$\mu$m fluxes brighter than 1\,mJy SMGs is 7$\times$ that of those
brighter than 4.2\,mJy \citep{Coppin06}.  Accounting for the errors in
the photometric redshifts of the ALESS SMGs, and applying this
correction to the number counts, in Fig.~\ref{fig:SFRd} we also show
the contribution to the comoving star formation rate density from
$S_{\rm 870\mu m}\gsim$\,1\,mJy SMGs.  This shows that SMGs contribute
$\sim $\,20\% of the total star formation over the redshift range
$z$\,=\,1--5.  Of course, we note that this estimate should be
considered a lower limits as we have not included ULIRGs which have
comparable luminosities as the SMGs, but with hotter than average dust
temperatures which makes them fainter at 870$\mu$m, dropping them below
the LESS flux limit.  The contribution to the star formation density
from these optical faint radio galaxies; OFRGs, \citep{Casey09} may
increase the contribution to the star formation rate density for ULIRGs
by a factor $\lsim $\,2$\times$ compared to the 870$\mu$m-only
selection we consider here.

Next, we compare these results to the semi-analytic galaxy formation
model from \citet{Baugh05} \citep[see also][]{Swinbank08,Gonzalez11}.
This model has the advantage that it is both cosmologically based, and
is required to fit the $z$\,=\,0 $K$-band luminosity function, {\it
  IRAS} 60$\mu$m number counts and galaxy bulge-to-disk ratios.  In
this model, the SMGs are dominated by bursts of star formation as a
result of major mergers, with the brightest SMGs ($S_{\rm 870\mu m}>
$\,5\,mJy) accounting for $\sim$1\% of the total star formation density
at $z$\,=\,2--4.  Integrating all sources with $S_{\rm 870\mu m}>
$\,1\,mJy this model also predicts that the fainter SMGs account for
$\sim$20--30\% of the total star formation density at these redshifts
\citep{Gonzalez11}.  Although the use of a top heavy ($x$\,=\,0) IMF in
the model complicates this comparison, Fig.~\ref{fig:SFRd} shows that
the {\sc galform} model appears to provide a reasonable description of
the ALESS data over the redshift range $z\sim $\,1--5.

Since the integral of the star formation history provides the total
stellar mass formed in a galaxy, we also measure the fraction of
stellar mass in SMGs compared to the total stellar mass density (as a
function of redshift).  We follow \citet{Guo12} and obtain the global
stellar mass density as a function of redshift by fitting a linear
relation to the evolution of the stellar mass density from Fig.~12 of
\citet{Marchesini10}.  Assuming the SMGs have a burst duration of
250\,Myr, we calculate the duty cycle as a function of redshift to infer the
total stellar mass formed by SMGs (we note that the burst
duration and duty cycle correction approximately trade off against each
other if the burst duration does not depend on redshift) and in
Fig.~\ref{fig:SFRd} we plot the fraction of stellar mass formed in SMGs
compared to the total stellar mass density as a function of redshift
for all ALESS SMGs with $S_{\rm 870}\geq $\,4.2\,mJy (again accounting
for the factor 2\,$\times$ under-density of the ECDFS).  In this plot,
we use the stellar masses for the ALESS SMGs derived by Simpson et
al.\ (2013), which have been estimated by integrating the
star formation histories and accounting for the mass loss due to winds
and supernovae.  This plot shows that at $z\sim $\,2--3, the bright
SMGs ($S_{\rm 870}\geq $\,4.2\,mJy) contribute 15\% of the total
stellar mass density at this epoch.  In contrast, by $z$\,=\,0 this
plot suggests that the total stellar mass formed in bright SMGs
comprises just $\sim $\,3--4\% of the total stellar mass density.  As
above, if we integrate the counts to $S_{\rm 870}\geq $\,1\,mJy, then
as Fig.~\ref{fig:SFRd} shows, SMGs account for 30--40\% of the total
stellar mass at $z\sim $\,2, and $\sim $\,15\% of the total stellar
mass density at $z$\,=\,0.

\subsection{Dust and Gas Mass Functions}

Exploiting the correlation between the dust and gas mass in local
galaxies, we can also use the mass ratio of the gas-to-dust to infer
the total H$_{2}$ mass in SMGs.  There has been considerable interest
in deriving the cold molecular gas masses in SMGs, since this provides
the raw ``fuel'' for star formation which determines the duration of
the starburst.  Most of the constraints on the gas masses have been
derived from low- and mid-$J$ $^{12}$CO observations and then have to
adopt CO--H$_2$ conversion factor
\citep[e.g.\ ][]{Frayer98,Greve05,Tacconi06,Tacconi08,Ivison11EVLA,Riechers11b,Bothwell13}.
In particular, \citet{Bothwell13} used $\sim $\,3\,Ms of low spatial
resolution observations with PdBI to derive the low-$J$ $^{12}$CO
properties of a sample 40 luminous SMGs (detecting 32 of them).
Constraining the molecular gas properties of a larger sample of SMGs,
even with ALMA, will therefore require a significant investment of
time.  An alternative approach to estimate the gas mass is to use the
(optically thin) continuum luminosity on the Rayligh-Jeans tail of the
dust SED to estimate the dust mass, and then use an appropriate
gas-to-dust ratio to derive the total mass of the molecular ISM.  A
detailed discussion of this technique and its application to both local
and high-redshift galaxies is given in \citet{Scoville13conf} review.

To estimate an appropriate gas-to-dust mass ratio ($\delta_{\rm GDR}$)
for SMGs, we exploit the CO-derived H$_{\rm 2}$ masses from
\citet{Bothwell13} who derive $M_{\rm
  H2}$\,=\,3.6\,$\pm$\,1.0\,$\times$\,10$^{10}$\,M$_{\odot}$ (including
non-detections).  Using far-infrared and radio photometry (from 70--870
and 1.4\,GHz), \citet{Magnelli12} fit the far-infrared SEDs of many of
the same galaxies to derive dust masses of $M_{\rm
  d}$\,=\,(3.9\,$\pm$\,0.5)\,$\times$\,10$^{8}$\,$M_{\odot}$,
suggesting an average gas-to-dust ratio of $\delta_{\rm
  GDR}$\,=\,90\,$\pm$\,25.  For comparison, we note that the Milky-Way
has a gas-to-dust ratio of $\delta_{\rm GDR}\sim $\,130;
\citealt{Jenkins04}, whilst the gas-to-dust ratio derived for the local
of star forming galaxies (across several galaxy types) from the SINGS
survey from \citet{Draine07} \citep[see also][]{Scoville13conf} is
$\delta_{\rm GDR}$\,=\,130\,$\pm$\,20 for galaxies with metallicities
above $Z$\,/\,$Z_{\sol}\gsim $\,0.2 (which is likely to represent a
lower-limit for SMGs; \citealt{Swinbank04,Takata06}).

It is also possible to derive a gas-to-dust ratio for the ALESS SMGs
using the stellar mass, star formation rate and metallicity ($Z$).  For
example, \citet{Maiolino08} suggest a mass-metallicity-star formation
rate plane of $Z$\,=\,8.90\,$+$\,0.47$\times x$ with
$x$\,=\,log(M$_{\star}$)\,$-$\,0.32\,log(SFR)\,$-$\,10.  The
gas-to-dust ratio can then be calculated by
$\delta_{GDR}$\,=\,10$^{-0.85\,Z\,+\,9.4}$ \citep{Magnelli12b}.  Using
the star formation rate and stellar mass for the ALESS SMGs in our
sample (Simpson et al.\ 2013), the average gas-to-dust ratio we derive
is $\delta_{GDR}$\,=\,75\,$\pm$\,10.  This is lower, but consistent
within the 1-$\sigma$ error of that derived from the (more direct)
$^{12}$CO and dust masses, although we caution that the significant
uncertainty in the stellar mass estimates for SMGs due to the
unconstrained star formation histories may dominate the difference in
$\delta_{\rm GDR}$ (Simpson et al.\ 2013).  For simplicity, in all of
the following analysis, we therefore adopt a single gas-to-dust ratio
of $\delta_{\rm GDR}$\,=\,90\,$\pm$\,25.

\begin{figure}
  \centerline{
    \psfig{file=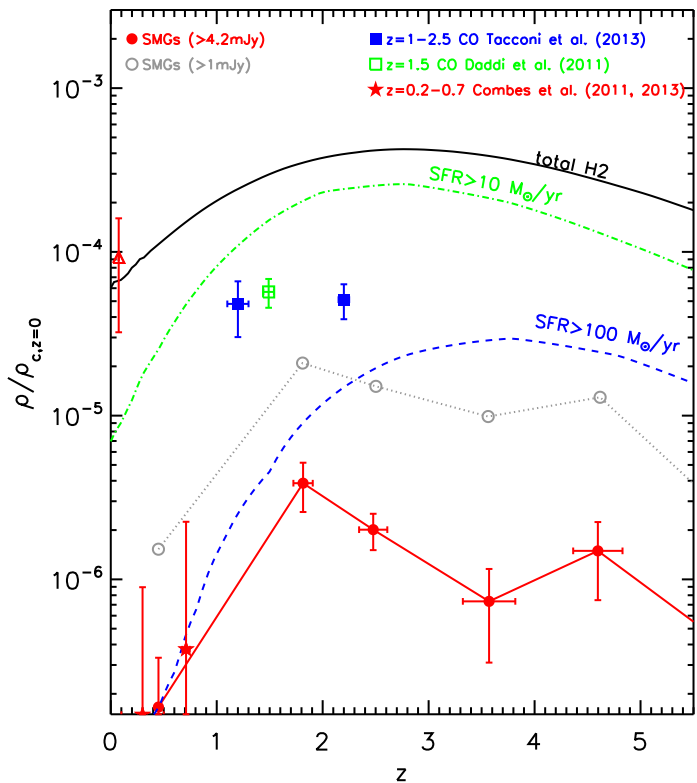,angle=0,width=3.5in}}
\caption{Global density of molecular hydrogen (normalised to the
  critical density at $z$\,=\,0), as a function of redshift for ALESS
  SMGs and other high-redshift starburst galaxies.  The solid line
  shows all ALESS SMGs with flux densities $S_{\rm 870\mu m}
  >$\,4.2\,mJy, whilst the dotted line shows the extrapolated
  contribution from all SMGs with flux densities $S_{ \rm 870\mu
    m}$\,$>$\,1\,mJy.  For comparison with measurements at $z$\,=\,0,
  we also overlay the data from \citet{Keres05}, as well as the model
  predictions for the evolution of the total gas density with redshift
  from \citet{Lagos11,Lagos13}.  At $S_{\rm 870\mu m} >$\,1\,mJy (which
  corresponds to a star formation rate of approximately $\sim
  $\,100\,M$_{\odot}$\,yr$^{-1}$ at $z\sim $\,2), the SMGs contribute
  only $\sim $\,5\% of the total $H_2$.  For comparisons with other
  high-redshift population, we also overlay the star-forming BX\,/\,BMs
  and BzK galaxies from \citet{Tacconi12} and luminous BzKs from
  \citet{Daddi10} which have comparable gas masses as the SMGs, but
  with space densities which are a factor $\sim $\,10$\times$ higher.
  These galaxies (which have typical star formation rates $\gsim
  $\,80\,M$_{\odot}$\,yr$^{-1}$) contain 10--15\% of the predicted
  total H$_{\rm 2}$ gas budget at $z\sim $\,1--2.  To provide a more
  detailed comparison of the contribution of galaxies with different
  star formation rates, we also overlay the model predictions from
  \citet{Lagos11} and \citet{Lagos13} for galaxies with star formation
  rates $> $\,100\,$M_{\odot}$\,yr$^{-1}$ (which provide a reasonable
  match to the $> $\,1\,mJy SMGs), and also for galaxies with star
  formation rates (SFR$>$\,=\,10\,$M_{\odot}$\,yr$^{-1}$).}
\label{fig:COz}
\end{figure}

Applying this gas-to-dust ratio to our estimates of the dust mass, this
suggests a median H$_2$ mass of $M_{\rm
  H2}$\,=\,(4.2\,$\pm$\,0.4)\,$\times$\,10$^{10}$\,M$_{\odot}$ for the
ALESS SMGs.  Together the average star formation rate and total H$_2$
mass of the SMGs suggest a gas depletion time-scale of $M_{\rm
  H_2}$\,/\,SFR$\sim $\,130\,$\pm$\,15\,Myr.  Assuming, on average that
the burst is observed half way through its lifetime and that the star
formation remains constant over the lifetime of the burst (with no
recycling of mass through supernovae or other mass loss), this suggests
a total burst duration in SMGs of $\sim $\,250\,Myr \citep[see
  also][]{Greve05,Hainline06,Tacconi06,Swinbank06b,Riechers11b,Ivison11EVLA,Hickox12,Bothwell13},
and a factor $\sim $\,3$\times$ longer than local ULIRGs of comparable
luminosity \citep[e.g.\ ][]{Solomon88,Gao04,Genzel10}.

In Fig.~\ref{fig:COLF} we plot the H$_2$ mass function for the bright
($S_{\rm 870}\geq $\,4.2\,mJy) ALESS SMGs (in units of $M_{\rm
  H_2}$\,/\,M$_{\odot}$ and Jy\,km\,s$^{-1}$\,Mpc$^{2}$).  
We also include on the plot the gas mass density of ``main-sequence''
starburst galaxies (BzKs and BX/BMs) from \citet{Daddi10} and
\citet{Tacconi10}.  Here, we have assumed that the six galaxies
observed by \citet{Daddi10} are representative of all star-forming BzK
galaxies (although we caution that the BzKs in Daddi et al.\ are the
most luminous sub-sample of the BzK population and so may be atypical).
Nevertheless, we adopt a space density of $\phi_{\rm
  BzK}$\,=\,(1.5\,$\pm$\,1.0)\,$\times$\,10$^{-4}$\,Mpc$^{-3}$ based on
the average BzK space density \citep{Daddi05}.  Recently,
\citet{Tacconi12} observed a much larger sample of massive,
main-sequence star-forming galaxies as part of the PHIBSS survey,
presenting detections of 52 star-forming galaxies between
$z$\,=\,1--1.5 and $z$\,=\,2.0--2.5.  These galaxies are selected from
a parent catalog of spectroscopically confirmed galaxies with star
formation rates $\gsim $\,30\,M$_{\odot}$\,yr$^{-1}$ and stellar masses
of M$_{\star} >$\,2.5\,$\times$\,10$^{11}$\,M$_{\odot}$ and have gas
masses of $M_{\rm
  H_2}$\,=\,(5.5\,$\pm$\,1.8)\,$\times$\,10$^{10}$\,M$_{\odot}$
(c.f. $M_{\rm
  H_2}$\,=\,4.2\,$\pm$\,0.4\,$\times$\,10$^{10}$\,M$_{\odot}$ for the
ALESS SMGs).  To estimate the space density of the parent population at
these limits, we use the \citet{Bower06} galaxy formation model (which
has been shown to provide a reasonable match to the high-redshift
stellar mass functions and star formation rates of galaxies).  For the
PHIBSS selection limits of SFR\,$\geq $\,30\,M$_{\odot}$\,yr$^{-1}$,
M$_{\star} >$\,2.5\,$\times$\,10$^{10}$\,M$_{\odot}$ and assuming a
median reddenning of A$_{\rm V}$\,=\,0.75 for these systems
\citep{ForsterSchreiber09}, the \citet{Bower06} model suggests space
densities for the parent populations of
$\phi_{z=1.0-1.5}$\,=\,1.4\,$\pm$\,0.6\,Mpc$^{-3}$ and
$\phi_{z=2.0-2.5}$\,=\,1.2\,$\pm$\,0.5\,Mpc$^{-3}$.  Thus, the space
densities of the PHIBSS galaxies and BzKs appear to be a factor $\sim
$\,8--10\,$\times$ higher than the ALESS SMGs, but with comparable gas
masses (Fig.~\ref{fig:COLF}).

In Fig.~\ref{fig:COz} we use the gas mass estimates for these samples
to investigate the total H$_2$ contained in star-forming galaxies as a
fraction of the total H$_2$ as a function of redshift.  As above, we
calculate the total gas density as a function of redshift contributed
by the bright ALESS SMGs ($S_{\rm 870\mu m}\geq $\,4.2\,mJy), and also
extrapolating to fainter limits ($S_{\rm 870\mu m}\geq $\,1\,mJy) using
the sub-mm counts from \citet{Coppin06}.  Of course, we caution that
extrapolating to $S_{\rm 870\mu m}> $\,1\,mJy assumes that the gas
masses scale with the 870$\mu$m flux, although the strong correlation
between CO, dust mass and far-infrared luminosity
\citep[e.g.\ ][]{Bothwell13} implies that this assumption is not
unreasonable.

To compare these results to other populations, we employ the $z$\,=\,0
data from \citet{Keres05}, who derived a total molecular gas content at
$z$\,=\,0 is $\rho$\,/\,$\rho_{\rm z=0}$\,=\,$0.9_{-0.6}^{+0.8}$\%
(where $\rho_{\rm z=0}$\,=\,3\,$H_0^2$\,/\,(8$\pi G$)).  For comparison
samples at intermediate redshift, we exploit the observations of 36
$z\sim $\,0.2--0.6 and 39 $z\sim $\,0.6--1.0 ULIRGs from
\citet{Combes11,Combes13} who derive gas masses from spectroscopy of 37
these galaxies using low-$J$ CO emission.  To estimate density of the
parent population of these intermediate redshift ULIRGs (which includes
optically bright and spectroscopically confirmed galaxies with $L_{\rm
  IR}>$\,2.8\,$\times$\,10$^{12}$\,$L_{\odot}$; above the 60$\mu$m {\it
  IRAS} {\it or} 70$\mu$m MIPS detection limits and $\delta
>$\,$-$12$^{\circ}$) using the the semi-analytic {\sc galform} model
from \citet{Baugh05}.  The space density of ULIRGs with these flux
limits should be $\sim $\,1.5\,$\times$\,10$^{-7}$\,Mpc$^{-3}$ and
$\sim $\,7.5\,$\times$\,10$^{-7}$\,Mpc$^{-3}$ for the $z\sim
$\,0.6--1.0 and $z\sim $\,0.2--0.6 populations respectively.
However, we caution that given the complex selection function and large
correction factors required in this calculation, we conservatively adopt
errors on the space density for this sample of at least a factor
4\,$\times$ at both redshifts.

We also plot the theoretical contribution of galaxies to the total
H$_2$ density as a function of total star formation rate and redshift
using the semi-analytic models of \citet{Lagos11,Lagos13} which is
based on the semi-analytic model of \citet{Baugh05} and
\citet{Bower06}.  As Fig.~\ref{fig:COz} shows, at $z\sim $\,2 the
bright ($\geq $\,4.2\,mJy) ALESS SMGs contain $\sim $\,1\% of the total
predicted H$_{\rm 2}$ density at $z$\,=\,1--3, although extrapolating
to galaxies with $S_{\rm 870\mu m}> $\,1\,mJy this rises to $\sim
$\,5\%.  In contrast, the star-forming BzKs and BX/BM galaxies (which
have typical star formation rates $\gsim $\,80\,M$_{\odot}$\,yr$^{-1}$)
contain 10--15\% of the predicted total H$_{\rm 2}$ gas budget at
$z\sim $\,1--2.  To provide a more detailed comparison of the
contribution of galaxies with different star formation rates, we also
overlay the predictions from the model for galaxies with star formation
rates $> $\,100\,$M_{\odot}$\,yr$^{-1}$ and
SFR$>$\,=\,10\,$M_{\odot}$\,yr$^{-1}$.  This model provides a
reasonable description of the data: model galaxies with star formation
rates SFR\,$>$\,100\,M$_{\odot}$\,yr$^{-1}$ (which is comparable to a
$S_{\rm 870\mu m}$ limit of 1\,mJy), should contribute $\sim $\,5--10\%
of the total at $z\sim $\,2, falling sharply to $<$0.1\% by $z$\,=\,1.

\section{Conclusions}

We have exploited the multi-wavelength imaging of the ECDFS to
investigate the far-infrared properties of a sample of 99
high-redshift, ALMA-detected sub-millimeter galaxies.  These galaxies
are precisely located from high-resolution ($\lsim $\,1.4$''$)
345\,GHz imaging, allowing us to measure the multi-wavelength
properties of the counterparts without recourse to statistical
associations.  Moreover, the sensitivity of the ALMA data also allow us
to derive the properties of SMGs to fainter sub-mm fluxes than
typically possible in single dish observations.  Our main findings are:

$\bullet$ Stacking the far-infrared imaging at the positions of the
ALESS SMGs, we show that their observed far-infrared SEDs peak close to
350-$\mu$m, as expected for a high-redshift galaxy population whose
dust temperatures are around $T_{\rm d}\sim $\,30\,K.  The SPIRE
colours of the individually radio-detected versus radio non-detected
subset of the ALESS SMGs are not statistically distinguishable.
However, when including the shorter wavelength PACS data, the SEDs of
the radio-detected SMGs appear to peak at shorter wavelengths compared
to those SMGs which are radio non-detected.  For a fixed characteristic
dust temperature, this is consistent with the radio-faint subset of the
ALESS SMGs lying at higher redshift, as also confirmed by their
photometric redshifts.

$\bullet$ By deblending the SPIRE 250, 350 and 500$\mu$m images of the
ECDFS based on a 24$\mu$m, radio and ALMA positional prior catalog we
find that 34 (out of 99) ALESS SMGs do not have a $>$3\,$\sigma$
counterpart at 250, 350 or 500$\mu$m.  Of these 34 galaxies, 30 are
also radio-undetected.  These SPIRE non-detections have a median
photometric redshift of $z$\,=\,3.3\,$\pm$\,0.5 which is higher than
the full ALESS SMG sample ($z$\,=\,2.5\,$\pm$\,0.2; Simpson et
al.\ 2013).  The median photometric redshift for ALESS SMGs which are
detected in at least two SPIRE bands and whose observed dust SEDs peak
at 250, 350 or 500$\mu$m are $z$\,=\,2.3\,$\pm$\,0.2, 2.5\,$\pm$\,0.3
and 3.5\,$\pm$\,0.5 respectively.

$\bullet$ We fit the far-infrared SEDs of the SMGs with a suite of dust
templates to derive the far-infrared luminosity and hence star
formation rate and characteristic dust temperature.  We derive a median
star formation rate for the SMGs in our sample of
SFR\,=\,330\,$\pm$\,30\,M$_{\odot}$\,yr$^{-1}$ with a range of
SFR\,=\,20--1030\,M$_{\odot}$\,yr$^{-1}$.  Concentrating on those ALESS
SMGs whose fluxes are brighter than $S_{\rm 870}\geq $\,4.2\,mJy (the
flux limit of the LESS survey; \citealt{Weiss09}), we derive a median
star formation rate of SFR\,=\,530\,$\pm$\,60\,M$_{\odot}$\,yr$^{-1}$.

$\bullet$ Accounting for the apparent under-density of bright SMGs in
ECDFS, we show that the contribution of ALESS SMGs with $S_{\rm 870\mu
  m}\geq $\,4.2\,mJy) to the co-moving star formation rate density
across the redshift range $z\sim $\,1--4 is $\lsim $\,1--2\% of the
total.  Integrating the 870$\mu$m counts down to 1\,mJy (the flux
corresponding to the luminosity where the contributions from the
far-infrared and UV to the bolometric output of galaxies typically
balance) then 870$\mu$m-selected SMGs should account for $\sim $\,20\%
of the total star formation across the same redshift range.

$\bullet$ By integrating the star formation histories of the SMGs
in our sample (and accounting for mass loss due to winds and
supernovae), we show that bright SMGs ($S_{\rm 870}\geq $\,4.2\,mJy)
contribute 15\% of the total stellar mass density at $z\sim $\,2.
Extrapolating to a flux limit of $S_{\rm 870}>$\,1\,mJy, SMGs account
for 30--40\% of the total stellar mass density at $z\sim $\,2 and $\sim
$\,15\% of the total stellar mass at $z$\,=\,0.

$\bullet$ Using the rest-frame 870-$\mu$m luminosities of the ALESS
SMGs, we infer an average dust masses of $M_{\rm
  d}$\,=\,(3.6\,$\pm$\,0.3)\,$\times$\,10$^{8}$\,M$_{\odot}$.  Adopting
a gas-to-dust ratio of $M_{\rm H_2}$\,/\,$M_{\rm d}$\,=\,90\,$\pm$\,25,
this suggests a typical cold gas mass of $M_{\rm H_2}\sim
$\,(4.2\,$\pm$\,0.4)\,$\times$\,10$^{10}$\,M$_{\odot}$.  Together the
average star formation rate and total H$_2$ mass of the SMGs suggest
gas depletion time-scales of $M_{\rm
  H_2}$\,/\,SFR\,=\,130\,$\pm$\,15\,Myr.

$\bullet$ Finally, we use our estimates of the H$_2$ mass to
investigate the contribution of star-forming galaxies to the cosmic
H$_2$ density as a function of redshift.  At $z\sim $\,2 the bright ($>
$\,4.2\,mJy) SMGs contain 1\% of the total H$_2$, although
extrapolating to  $S_{\rm 870\mu m}> $\,1mJy this rises to $\sim $\,5\%.
We show that this is consistent with the latest theoretical models
which predict that galaxies with star formation rates
SFR\,$>$\,100\,M$_{\odot}$\,yr$^{-1}$, should contribute $\sim
$\,5--10\% of the total at $z\sim $\,2, falling sharply to $<$0.1\% by
$z$\,=\,1.

We have presented an analysis of the far-infrared and radio properties
of an unbiased sample of 870$\mu$m-selected SMGs in the ECDFS whose
positions have been precisely measured with ALMA.  We show that the
SMGs in our sample have typical star formation rates of
SFR\,=\,310\,$\pm$\,30\,$M_{\odot}$\,yr$^{-1}$ and by integrating the
counts to a flux limit of $S_{\rm 870}\geq $\,1\,mJy, we show that SMGs
can account for $\sim $\,20\% of the co-moving star formation density
at $z\sim $\,1--5 and estimate that these systems contain $\sim $\,10\%
of the total molecular gas budget at this epoch.  In a future paper, we
will combine high resolution \emph{HST} and ALMA imaging of ALESS SMGs
with measurements of their internal dynamics to investigate how mergers
and interactions trigger the high star formation rates seen in these
systems.  Such observations will allow us to better estimate the
distribution and intensity of star formation and the time-scales
involved in the encounters that appear to drive the rapid star
formation.

\section*{acknowledgments}

We would like to thank the referee, Catlin Casey, for a constructive
report which improved the content and clarity of this paper.  The ALMA
observations were carried out under program 2011.0.00294.S.  ALMA is a
partnership of ESO (representing its member states), NSF (USA) and NINS
(Japan), together with NRC (Canada) and NSC and ASIAA (Taiwan), in
cooperation with the Republic of Chile. The Joint ALMA Observatory is
operated by ESO, AUI/NRAO and NAOJ.  This publication is also partly
based on data acquired with the APEX under programme IDs 078.F-9028(A),
079.F-9500(A), 080.A-3023(A) and 081.F-9500(A). APEX is a collaboration
between the Max-Planck-Institut fur Radioastronomie, the European
Southern Observatory and the Onsala Space Observatory.  We also make
use of data from our ESO/VLT large program ($z$LESS) PID: 183.A-0666
and VLT/XSHOOTER program 090.A.0927.  This research also made use of
data taken as part of the HerMES Key Programme from the SPIRE
instrument team, ESAC scientists and a mission scientist.
\emph{Herschel} is an ESA space observatory with science instruments
provided by European-led Principal Investigator consortia and with
important participation from NASA.  All of the ALMA, \emph{Herschel},
VLA and {\it Spitzer} data employed in this analysis are available
through the ESO, Herschel VLA and {\it Spitzer} archives.  AMS
gratefully acknowledges an STFC Advanced Fellowship through grant
number ST/H005234/1.  IRS also acknowledges STFC grant ST/I001573/1, a
Leverhume Fellowship, the ERC Advanced Grant programme {\sc dustygal}
and a Royal Society Wolfson Merit Award.  ALRD acknowledges an STFC
studentship (ST/F007299/1).  AK acknowledge support by the
Collaborative Research Council 956, sub-project A1, funded by the
Deutsche Forschungsgemeinschaft (DFG).  JLW acknowledges support from
Program HST-GO-12866.13-A which was provided by NASA through a grant
from the Space Telescope Science Institute, which is operated by the
Association of Universities for Research in Astronomy, Incorporated,
under NASA contract NAS5-26555 and the Dark Cosmology Centre which is
funded by the Danish National Research Foundation.


\appendix
\section{Example SPIRE images, models and residuals for LESS regions}
\begin{landscape}

%
%
\begin{figure*}
\hspace{-4cm}  \centerline{\psfig{file=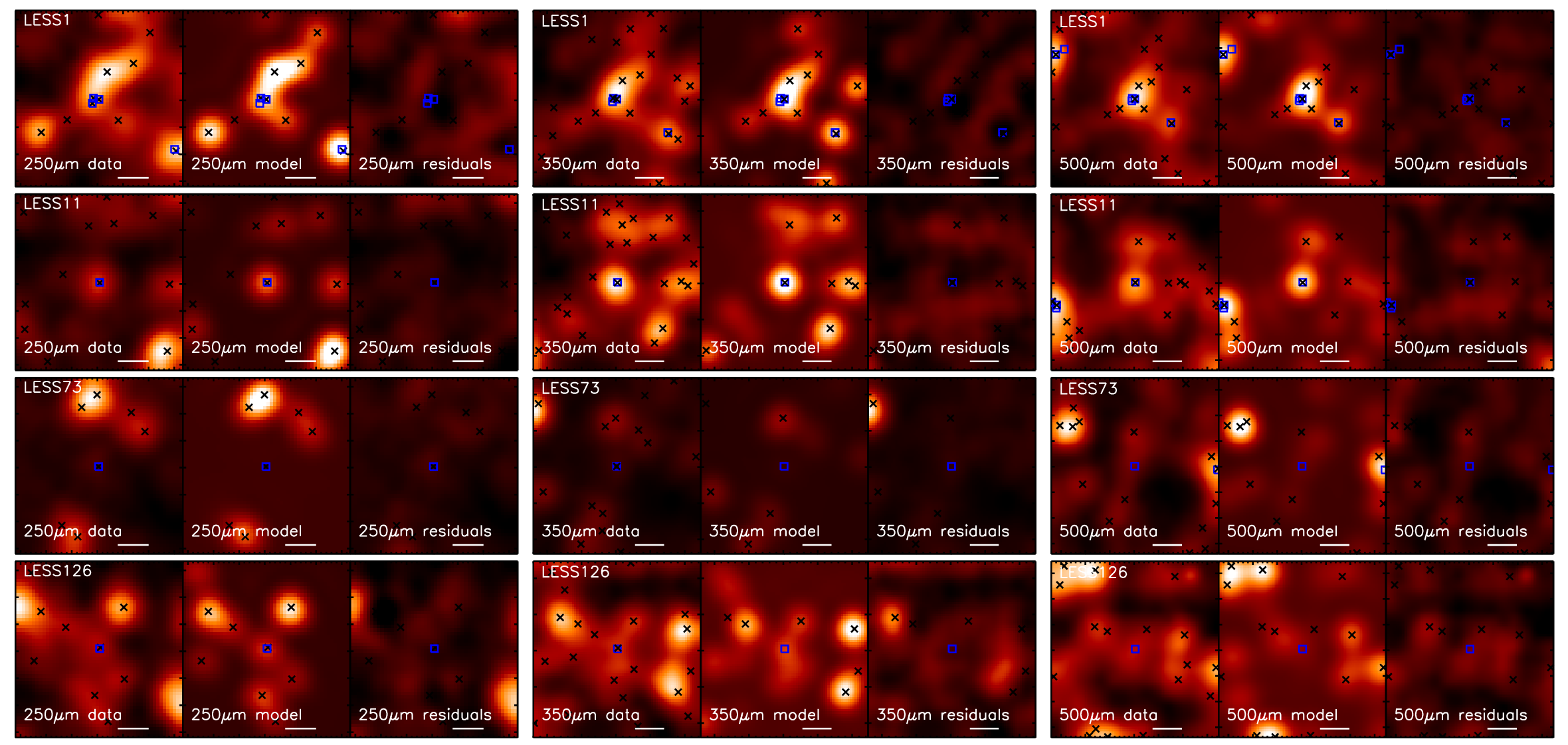,angle=90,width=9in}}
\caption{Example 250, 350 \& 500$\mu$m images, models and residuals for
  the ALMA SMGs in our sample.  These postage stamps show the images
  for the four ALMA SMGs whose SEDs are shown in Fig.~\ref{fig:SEDs}.
  For each image, the data, model and residuals are on the same colour
  scale.  In the ``data'' panel, black crosses denote the galaxies in
  the positional prior catalog for that wavelength (only galaxies detected at
  $>$2\,$\sigma$ at 250$\mu$m are used in deblending the 350$\mu$m and
  similarly for 500$\mu$m).  To highlight this, in the ``model'' and
  ``residuals'' panels, black crosses denote those galaxies which are
  detected at that wavelength above the flux limits (see \S3).  In all
  panels, we mark the positions of the ALESS SMGs as squares.  The 250,
  350 and 500$\mu$m images are 1.7, 2.5 and 3.4 arc-minutes across
  respectively ($\sim $\,6 beams in each case).}
\label{fig:DataModelResiduals}
\end{figure*}
\end{landscape}

In Fig.~\ref{fig:DataModelResiduals} we show the SPIRE 250, 350 and
500-$\mu$m images, models and residual maps for four LESS SMG regions
whose far-infrared SEDs are shown in Fig.~\ref{fig:SEDs}.

In Table~A1 we provide the ID, far-infrared and radio fluxes for the 96
ALESS SMGs which have photometric redshifts in our sample.  We also
list their derived properties, dust mass, far-infrared luminosity and
characteristic dust temperature.

\clearpage
  

%
%
\begin{table*}
{\tiny
\caption{Far-Infrared and radio flux densities of ALESS SMGs and their derived properties}
\begin{tabular}{lcccccccccc}
\hline
\hline
ID          & S$_{870\mu m}$ &  S$_{\rm 1.4GHz}$ & S$_{500\mu m}$ & S$_{350\mu m}$ & S$_{250\mu m}$ & S$_{\rm 24\mu m}$     & $z_{\rm phot}$   & $\log$($M_{\rm dust}$) & $L_{\rm 8-1000\mu m}$ & $T_{\rm d^{*}}$ \\
            & (mJy)         & ($\mu$Jy)       & (mJy)        & (mJy)         &  (mJy)        & (mJy)               &                &                            & ($\times$\,10$^{12}$\,$L_{\odot}$) & (K) \\
\hline
LESS1.1 & 6.75\,$\pm$\,0.49 & $<$\,22.8 & $<$\,16.65 & $<$\,15.59 & $<$\,8.30 & $<$\,45.0 &                                                               4.34$_{-1.43}^{+2.66}$ & 9.13\,$\pm$\,0.03 & 3.3$_{-1.6}^{+7.7}$ & 36[24]                  \\    
LESS1.2 & 3.48\,$\pm$\,0.43 & $<$\,22.8 & $<$\,16.65 & $<$\,15.59 & $<$\,9.14 & $<$\,45.0 &                                                               4.65$_{-1.02}^{+2.34}$ & 8.86\,$\pm$\,0.05 & 3.4$_{-1.6}^{+2.9}$ & 44[32]                  \\    
LESS1.3 & 1.89\,$\pm$\,0.42 & $<$\,22.8 & $<$\,17.62 & 16.69\,$\pm$\,3.57 & 10.70\,$\pm$\,2.31 & $<$\,45.0 &                                              2.85$_{-0.30}^{+0.20}$ & 8.25\,$\pm$\,0.09 & 3.2$_{-1.8}^{+1.2}$ & 31[40]                  \\    
LESS2.1 & 3.81\,$\pm$\,0.42 & 236.90\,$\pm$\,7.90 & 19.90\,$\pm$\,4.29 & 17.78\,$\pm$\,3.76 & 17.66\,$\pm$\,3.21 & $<$\,45.0 &                            1.96$_{-0.20}^{+0.27}$ & 8.39\,$\pm$\,0.05 & 4.4$_{-1.2}^{+0.0}$ & 24[40]                  \\    
LESS2.2 & 4.23\,$\pm$\,0.67 & $<$\,22.8 & 11.53\,$\pm$\,2.93 & 17.62\,$\pm$\,3.71 & 11.58\,$\pm$\,2.46 & $<$\,45.0 &                                      3.92$_{-1.42}^{+0.48}$ & 8.68\,$\pm$\,0.06 & 5.6$_{-3.2}^{+2.1}$ & 42[37]                  \\    
LESS3.1 & 8.28\,$\pm$\,0.40 & $<$\,22.8 & 25.80\,$\pm$\,5.03 & 23.33\,$\pm$\,4.29 & 15.68\,$\pm$\,2.93 & $<$\,45.0 &                                      3.90$_{-0.59}^{+0.50}$ & 8.98\,$\pm$\,0.02 & 9.0$_{-3.0}^{+4.7}$ & 36[35]                  \\    
LESS5.1 & 7.78\,$\pm$\,0.68 & 41.70\,$\pm$\,8.80 & 22.47\,$\pm$\,4.61 & 22.90\,$\pm$\,4.25 & 10.57\,$\pm$\,2.33 & $<$\,45.0 &                             2.86$_{-0.04}^{+0.05}$ & 8.54\,$\pm$\,0.04 & 4.9$_{-0.7}^{+0.4}$ & 27[38]                  \\    
LESS6.1 & 5.98\,$\pm$\,0.41 & 48.90\,$\pm$\,8.10 & 25.29\,$\pm$\,4.95 & 20.45\,$\pm$\,3.99 & 14.72\,$\pm$\,2.83 & 89.65\,$\pm$\,15.00 &                   0.45$_{-0.04}^{+0.06}$ & 8.25\,$\pm$\,0.03 & 0.2$_{-0.0}^{+0.1}$ & 11[19]                  \\    
LESS7.1 & 6.10\,$\pm$\,0.32 & 78.60\,$\pm$\,7.10 & 34.95\,$\pm$\,6.12 & 43.60\,$\pm$\,5.74 & 32.16\,$\pm$\,4.54 & 494.70\,$\pm$\,15.00 &                  2.50$_{-0.16}^{+0.12}$ & 8.99\,$\pm$\,0.02 & 6.4$_{-0.9}^{+1.2}$ & 31[32]                  \\    
LESS9.1 & 8.75\,$\pm$\,0.47 & 34.80\,$\pm$\,7.10 & 26.60\,$\pm$\,5.15 & 17.43\,$\pm$\,3.66 & 14.50\,$\pm$\,2.78 & $<$\,45.0 &                             4.50$_{-2.33}^{+0.54}$ & 9.08\,$\pm$\,0.02 & 8.4$_{-3.5}^{+7.1}$ & 38[32]                  \\    
LESS10.1 & 5.25\,$\pm$\,0.50 & 98.70\,$\pm$\,6.30 & $<$\,18.04 & 19.01\,$\pm$\,3.87 & 21.61\,$\pm$\,3.58 & 133.75\,$\pm$\,15.00 &                         2.02$_{-0.09}^{+0.09}$ & 8.42\,$\pm$\,0.04 & 3.1$_{-0.3}^{+0.1}$ & 26[38]                  \\   
LESS11.1 & 7.29\,$\pm$\,0.41 & 55.40\,$\pm$\,6.90 & 23.82\,$\pm$\,4.75 & 23.04\,$\pm$\,4.23 & 15.12\,$\pm$\,2.85 & 108.16\,$\pm$\,15.00 &                 2.83$_{-0.50}^{+1.88}$ & 8.84\,$\pm$\,0.02 & 7.9$_{-1.9}^{+0.7}$ & 29[33]                  \\   
LESS13.1 & 8.01\,$\pm$\,0.59 & 24.80\,$\pm$\,6.30 & 11.73\,$\pm$\,2.90 & 11.77\,$\pm$\,2.89 & 8.09\,$\pm$\,1.90 & $<$\,45.0 &                             3.25$_{-0.46}^{+0.64}$ & 8.92\,$\pm$\,0.03 & 5.4$_{-1.7}^{+2.3}$ & 26[24]                  \\   
LESS14.1 & 7.47\,$\pm$\,0.52 & 91.20\,$\pm$\,8.10 & 29.99\,$\pm$\,5.53 & 27.63\,$\pm$\,4.64 & 21.35\,$\pm$\,3.52 & 67.35\,$\pm$\,15.00 &                  4.47$_{-0.88}^{+2.54}$ & 9.15\,$\pm$\,0.03 & 9.4$_{-0.3}^{+0.6}$ & 42[33]                  \\   
LESS15.1 & 9.01\,$\pm$\,0.37 & 32.10\,$\pm$\,9.70 & 21.11\,$\pm$\,4.39 & 23.57\,$\pm$\,4.29 & 12.62\,$\pm$\,2.58 & 186.09\,$\pm$\,15.00 &                 1.93$_{-0.33}^{+0.62}$ & 9.51\,$\pm$\,0.02 & 1.3$_{-0.2}^{+2.7}$ & 21[19]                  \\   
LESS15.3 & 1.95\,$\pm$\,0.52 & $<$\,22.8 & $<$\,15.93 & $<$\,15.33 & $<$\,10.14 & $<$\,45.0 &                                                             3.15$_{-0.65}^{+0.65}$ & 8.65\,$\pm$\,0.10 & 1.5$_{-0.5}^{+2.8}$ & 35[19]                  \\   
LESS17.1 & 8.44\,$\pm$\,0.46 & 122.20\,$\pm$\,7.00 & 27.57\,$\pm$\,5.24 & 32.58\,$\pm$\,5.04 & 23.60\,$\pm$\,3.78 & 194.86\,$\pm$\,15.00 &                1.51$_{-0.07}^{+0.10}$ & 9.38\,$\pm$\,0.02 & 1.6$_{-0.1}^{+0.5}$ & 20[19]                  \\   
LESS18.1 & 4.38\,$\pm$\,0.54 & 129.40\,$\pm$\,6.90 & 32.24\,$\pm$\,5.81 & 40.96\,$\pm$\,5.60 & 41.70\,$\pm$\,5.30 & 1020.42\,$\pm$\,15.00 &               2.04$_{-0.06}^{+0.10}$ & 8.79\,$\pm$\,0.05 & 5.4$_{-0.8}^{+0.8}$ & 30[33]                  \\   
LESS19.1 & 4.98\,$\pm$\,0.42 & 34.70\,$\pm$\,6.90 & 16.02\,$\pm$\,3.63 & 24.85\,$\pm$\,4.41 & 19.90\,$\pm$\,3.38 & $<$\,45.0 &                            2.41$_{-0.11}^{+0.17}$ & 8.41\,$\pm$\,0.04 & 3.6$_{-0.5}^{+0.4}$ & 30[38]                  \\   
LESS19.2 & 1.98\,$\pm$\,0.47 & $<$\,22.8 & $<$\,17.62 & $<$\,16.50 & $<$\,13.50 & $<$\,45.0 &                                                             2.17$_{-0.10}^{+0.09}$ & 8.01\,$\pm$\,0.09 & 1.5$_{-0.5}^{+1.3}$ & 30[38]                  \\   
LESS22.1 & 4.48\,$\pm$\,0.54 & 65.20\,$\pm$\,8.60 & 23.75\,$\pm$\,4.74 & 32.70\,$\pm$\,5.05 & 25.03\,$\pm$\,3.88 & 465.14\,$\pm$\,15.00 &                 1.88$_{-0.23}^{+0.18}$ & 9.08\,$\pm$\,0.05 & 2.7$_{-0.4}^{+2.5}$ & 27[24]                  \\   
LESS23.1 & 6.74\,$\pm$\,0.37 & $<$\,22.8 & $<$\,16.65 & 16.20\,$\pm$\,3.51 & 13.71\,$\pm$\,2.72 & $<$\,45.0 &                                             4.99$_{-2.55}^{+2.01}$ & 9.01\,$\pm$\,0.02 & 6.4$_{-2.7}^{+4.9}$ & 43[35]                  \\   
LESS23.7 & 1.76\,$\pm$\,0.49 & $<$\,22.8 & $<$\,17.07 & $<$\,16.49 & $<$\,16.64 & $<$\,45.0 &                                                             2.90$_{-0.40}^{+1.20}$ & 8.69\,$\pm$\,0.11 & 3.0$_{-2.0}^{+1.6}$ & 40[19]                  \\   
LESS25.1 & 6.21\,$\pm$\,0.47 & 63.40\,$\pm$\,7.60 & 19.40\,$\pm$\,4.13 & 26.13\,$\pm$\,4.52 & 28.01\,$\pm$\,4.19 & 164.17\,$\pm$\,15.00 &                 2.24$_{-0.17}^{+0.07}$ & 8.43\,$\pm$\,0.03 & 7.5$_{-1.2}^{+0.4}$ & 29[41]                  \\   
LESS29.1 & 5.90\,$\pm$\,0.43 & 49.80\,$\pm$\,9.60 & 23.96\,$\pm$\,4.77 & 25.57\,$\pm$\,4.47 & 21.71\,$\pm$\,3.56 & $<$\,45.0 &                            2.66$_{-0.76}^{+2.94}$ & 8.47\,$\pm$\,0.03 & 10.0$_{-5.4}^{+0.9}$& 31[41]                  \\  
LESS31.1 & 8.12\,$\pm$\,0.37 & 30.40\,$\pm$\,6.90 & 15.12\,$\pm$\,3.46 & 18.85\,$\pm$\,3.79 & 11.06\,$\pm$\,2.34 & $<$\,45.0 &                            2.89$_{-0.41}^{+1.80}$ & 8.68\,$\pm$\,0.02 & 7.3$_{-2.9}^{+5.3}$ & 26[32]                  \\   
LESS37.1 & 2.92\,$\pm$\,0.41 & 31.50\,$\pm$\,8.40 & 12.28\,$\pm$\,3.00 & 18.99\,$\pm$\,3.81 & 18.34\,$\pm$\,3.22 & 258.49\,$\pm$\,15.00 &                 3.53$_{-0.31}^{+0.56}$ & 8.87\,$\pm$\,0.06 & 4.3$_{-0.5}^{+0.3}$ & 44[32]               \\   
LESS37.2 & 1.65\,$\pm$\,0.44 & $<$\,22.8 & $<$\,14.00 & $<$\,13.97 & $<$\,11.80 & $<$\,45.0 &                                                             4.87$_{-0.40}^{+0.21}$ & 8.52\,$\pm$\,0.10 & 1.9$_{-1.2}^{+2.1}$ & 55[32]               \\   
LESS39.1 & 4.33\,$\pm$\,0.34 & 49.80\,$\pm$\,8.10 & 15.01\,$\pm$\,3.45 & 19.04\,$\pm$\,3.81 & 8.44\,$\pm$\,2.01 & 122.49\,$\pm$\,15.00 &                  2.44$_{-0.23}^{+0.17}$ & 8.70\,$\pm$\,0.03 & 3.8$_{-0.8}^{+0.4}$ & 27[32]               \\   
LESS41.1 & 4.88\,$\pm$\,0.61 & $<$\,22.8 & 16.76\,$\pm$\,3.73 & 22.84\,$\pm$\,4.21 & 15.57\,$\pm$\,2.92 & 340.47\,$\pm$\,15.00 &                          2.75$_{-0.72}^{+4.25}$ & 8.68\,$\pm$\,0.05 & 5.3$_{-2.7}^{+0.4}$ & 31[35]               \\   
LESS41.3 & 2.68\,$\pm$\,0.75 & $<$\,22.8 & $<$\,13.42 & $<$\,15.04 & $<$\,10.14 & $<$\,45.0 &                                                             3.10$_{-0.60}^{+1.30}$ & 8.36\,$\pm$\,0.11 & 1.9$_{-0.8}^{+3.4}$ & 33[32]               \\   
LESS43.1 & 2.30\,$\pm$\,0.42 & $<$\,22.8 & 12.30\,$\pm$\,3.04 & 14.71\,$\pm$\,3.34 & 9.40\,$\pm$\,2.17 & 218.09\,$\pm$\,15.00 &                           1.71$_{-0.12}^{+0.20}$ & 8.79\,$\pm$\,0.07 & 1.0$_{-0.2}^{+0.7}$ & 23[23]               \\   
LESS45.1 & 6.03\,$\pm$\,0.54 & 35.50\,$\pm$\,6.70 & 18.22\,$\pm$\,3.95 & 20.51\,$\pm$\,3.96 & 12.94\,$\pm$\,2.57 & 66.93\,$\pm$\,15.00 &                  2.34$_{-0.67}^{+0.26}$ & 8.38\,$\pm$\,0.04 & 3.5$_{-0.5}^{+0.2}$ & 26[38]               \\   
LESS49.1 & 6.00\,$\pm$\,0.68 & 84.50\,$\pm$\,8.40 & $<$\,15.41 & 20.88\,$\pm$\,4.08 & 17.65\,$\pm$\,3.25 & 133.14\,$\pm$\,15.00 &                         2.76$_{-0.14}^{+0.11}$ & 8.73\,$\pm$\,0.05 & 7.3$_{-2.3}^{+0.8}$ & 32[32]               \\   
LESS49.2 & 1.80\,$\pm$\,0.46 & 37.00\,$\pm$\,8.20 & $<$\,14.29 & 9.28\,$\pm$\,2.53 & 18.70\,$\pm$\,3.29 & $<$\,45.0 &                                     1.47$_{-0.10}^{+0.07}$ & 7.94\,$\pm$\,0.10 & 0.9$_{-0.2}^{+0.7}$ & 23[38]               \\   
LESS51.1 & 4.70\,$\pm$\,0.39 & 34.50\,$\pm$\,7.20 & $<$\,14.86 & $<$\,15.33 & $<$\,12.37 & 231.94\,$\pm$\,15.00 &                                         1.22$_{-0.06}^{+0.03}$ & 8.96\,$\pm$\,0.03 & 0.9$_{-0.3}^{+0.3}$ & 17[19]               \\   
LESS55.1 & 3.99\,$\pm$\,0.36 & $<$\,22.8 & $<$\,10.60 & $<$\,8.00 & $<$\,7.00 & $<$\,45.0 &                                                               2.05$_{-0.13}^{+0.15}$ & 8.90\,$\pm$\,0.04 & 0.5$_{-0.1}^{+0.0}$ & 21[19]               \\   
LESS55.2 & 2.35\,$\pm$\,0.60 & $<$\,22.8 & $<$\,10.60 & $<$\,8.00 & $<$\,7.00 & $<$\,45.0 &                                                               4.20$_{-0.90}^{+0.50}$ & 8.96\,$\pm$\,0.10 & 2.4$_{-1.3}^{+1.6}$ & 40[19]               \\   
LESS55.5 & 1.37\,$\pm$\,0.37 & $<$\,22.8 & $<$\,10.60 & $<$\,8.00 & $<$\,7.00 & $<$\,45.0 &                                                               2.35$_{-0.13}^{+0.11}$ & 7.89\,$\pm$\,0.10 & 0.9$_{-0.4}^{+1.2}$ & 28[38]               \\   
LESS57.1 & 3.56\,$\pm$\,0.61 & 51.00\,$\pm$\,7.20 & 20.71\,$\pm$\,4.32 & 15.31\,$\pm$\,3.41 & 15.85\,$\pm$\,2.97 & 283.72\,$\pm$\,15.00 &                 2.95$_{-0.10}^{+0.05}$ & 8.86\,$\pm$\,0.07 & 4.4$_{-0.7}^{+0.8}$ & 32[32]               \\   
LESS59.2 & 1.94\,$\pm$\,0.44 & $<$\,22.8 & 11.24\,$\pm$\,2.83 & 14.42\,$\pm$\,3.28 & 9.89\,$\pm$\,2.21 & $<$\,45.0 &                                      2.09$_{-0.29}^{+0.78}$ & 8.02\,$\pm$\,0.09 & 1.6$_{-0.3}^{+1.8}$ & 27[38]               \\   
LESS61.1 & 4.29\,$\pm$\,0.51 & $<$\,22.8 & $<$\,10.60 & $<$\,9.94 & 7.06\,$\pm$\,1.76 & $<$\,45.0 &                                                       6.52$_{-0.34}^{+0.36}$ & 8.65\,$\pm$\,0.05 & 5.7$_{-2.2}^{+2.6}$ & 52[37]               \\   
LESS63.1 & 5.59\,$\pm$\,0.35 & $<$\,22.8 & $<$\,13.71 & $<$\,13.56 & 7.44\,$\pm$\,1.79 & $<$\,45.0 &                                                      1.87$_{-0.33}^{+0.10}$ & 9.19\,$\pm$\,0.03 & 0.6$_{-0.0}^{+0.1}$ & 20[19]               \\   
LESS65.1 & 4.16\,$\pm$\,0.43 & $<$\,22.8 & $<$\,11.79 & $<$\,9.15 & $<$\,7.00 & $<$\,45.0 &                                                               2.82$_{-0.36}^{+0.95}$ & 8.47\,$\pm$\,0.04 & 3.9$_{-1.7}^{+1.8}$ & 27[32]               \\   
LESS66.1 & 2.50\,$\pm$\,0.48 & 69.80\,$\pm$\,8.10 & $<$\,10.60 & 16.33\,$\pm$\,3.49 & 20.07\,$\pm$\,3.42 & 576.23\,$\pm$\,15.00 &                         2.33$_{-0.04}^{+0.05}$ & 8.67\,$\pm$\,0.08 & 3.2$_{-0.6}^{+0.4}$ & 35[33]               \\   
LESS67.1 & 4.50\,$\pm$\,0.38 & 73.90\,$\pm$\,6.90 & 22.99\,$\pm$\,4.63 & 35.68\,$\pm$\,5.26 & 33.68\,$\pm$\,4.61 & 732.63\,$\pm$\,15.00 &                 2.14$_{-0.09}^{+0.05}$ & 8.85\,$\pm$\,0.04 & 5.3$_{-1.3}^{+0.7}$ & 31[32]               \\   
LESS67.2 & 1.73\,$\pm$\,0.41 & $<$\,22.8 & $<$\,14.86 & $<$\,16.37 & 7.22\,$\pm$\,1.87 & $<$\,45.0 &                                                      2.05$_{-0.16}^{+0.06}$ & 7.88\,$\pm$\,0.09 & 1.1$_{-0.5}^{+0.3}$ & 24[38]               \\   
LESS68.1 & 3.70\,$\pm$\,0.56 & $<$\,22.8 & $<$\,12.58 & 8.07\,$\pm$\,2.31 & 8.38\,$\pm$\,1.98 & $<$\,45.0 &                                               3.60$_{-1.10}^{+1.10}$ & 8.58\,$\pm$\,0.06 & 3.4$_{-1.1}^{+1.7}$ & 28[32]               \\   
LESS69.1 & 4.85\,$\pm$\,0.63 & $<$\,22.8 & $<$\,13.14 & $<$\,13.97 & $<$\,9.14 & 87.91\,$\pm$\,15.00 &                                                    2.34$_{-0.44}^{+0.27}$ & 8.63\,$\pm$\,0.05 & 2.2$_{-1.5}^{+0.3}$ & 25[24]               \\   
LESS69.2 & 2.36\,$\pm$\,0.56 & $<$\,22.8 & $<$\,13.42 & $<$\,13.97 & $<$\,9.62 & $<$\,45.0 &                                                              4.75$_{-1.05}^{+0.35}$ & 8.71\,$\pm$\,0.09 & 2.6$_{-1.0}^{+2.5}$ & 48[35]               \\   
LESS69.3 & 2.05\,$\pm$\,0.56 & $<$\,22.8 & $<$\,13.14 & $<$\,12.68 & $<$\,7.00 & $<$\,45.0 &                                                              4.80$_{-1.10}^{+0.30}$ & 9.02\,$\pm$\,0.10 & 2.3$_{-0.9}^{+2.1}$ & 45[19]               \\   
LESS70.1 & 5.23\,$\pm$\,0.45 & 325.40\,$\pm$\,7.60 & 24.09\,$\pm$\,4.78 & 33.89\,$\pm$\,5.13 & 33.25\,$\pm$\,4.61 & 436.16\,$\pm$\,15.00 &                2.28$_{-0.06}^{+0.05}$ & 8.78\,$\pm$\,0.04 & 7.9$_{-1.2}^{+0.8}$ & 31[33]               \\   
LESS71.1 & 2.85\,$\pm$\,0.60 & 199.00\,$\pm$\,8.70 & 24.41\,$\pm$\,4.83 & 49.13\,$\pm$\,6.02 & 51.97\,$\pm$\,5.98 & 609.47\,$\pm$\,15.00 &                2.48$_{-0.11}^{+0.21}$ & 8.62\,$\pm$\,0.08 & 14.3$_{-0.9}^{+0.0}$& 38[52]               \\  
LESS71.3 & 1.36\,$\pm$\,0.38 & $<$\,22.8 & $<$\,14.00 & $<$\,16.42 & $<$\,17.82 & $<$\,45.0 &                                                             2.73$_{-0.25}^{+0.22}$ & 8.22\,$\pm$\,0.11 & 2.7$_{-1.7}^{+1.0}$ & 42[41]               \\   
LESS72.1 & 4.91\,$\pm$\,0.50 & $<$\,22.8 & $<$\,13.14 & $<$\,15.33 & $<$\,11.80 & $<$\,45.0 &                                                             4.15$_{-1.65}^{+0.55}$ & 9.13\,$\pm$\,0.04 & 4.1$_{-2.6}^{+4.1}$ & 40[19]               \\   
LESS73.1 & 6.09\,$\pm$\,0.47 & 24.00\,$\pm$\,6.30 & $<$\,10.60 & $<$\,8.00 & $<$\,7.00 & $<$\,45.0 &                                                      5.18$_{-0.45}^{+0.43}$ & 8.97\,$\pm$\,0.03 & 5.6$_{-1.1}^{+1.8}$ & 38[32]               \\   
LESS74.1 & 4.64\,$\pm$\,0.69 & 48.00\,$\pm$\,8.20 & 12.52\,$\pm$\,3.06 & 21.47\,$\pm$\,4.12 & 20.57\,$\pm$\,3.45 & 164.91\,$\pm$\,15.00 &                 1.80$_{-0.13}^{+0.13}$ & 8.27\,$\pm$\,0.06 & 2.8$_{-1.0}^{+0.0}$ & 27[38]               \\   
LESS75.1 & 3.17\,$\pm$\,0.45 & 74.90\,$\pm$\,8.50 & 25.68\,$\pm$\,5.01 & 29.18\,$\pm$\,4.80 & 23.99\,$\pm$\,3.84 & 1240.04\,$\pm$\,15.00 &                2.39$_{-0.06}^{+0.08}$ & 8.86\,$\pm$\,0.06 & 4.2$_{-0.4}^{+0.9}$ & 31[32]               \\   
LESS75.4 & 1.30\,$\pm$\,0.37 & $<$\,22.8 & $<$\,16.87 & $<$\,14.71 & $<$\,10.68 & $<$\,45.0 &                                                             2.10$_{-0.34}^{+0.29}$ & 7.89\,$\pm$\,0.11 & 1.1$_{-0.6}^{+1.5}$ & 29[38]               \\   
LESS76.1 & 6.42\,$\pm$\,0.58 & 45.40\,$\pm$\,9.50 & $<$\,12.04 & $<$\,9.53 & $<$\,7.00 & $<$\,45.0 &                                                      4.50$_{-2.00}^{+0.20}$ & 8.83\,$\pm$\,0.04 & 6.8$_{-1.9}^{+1.1}$ & 35[33]               \\   
LESS79.1 & 4.12\,$\pm$\,0.37 & $<$\,22.8 & 17.89\,$\pm$\,3.92 & 16.11\,$\pm$\,3.52 & 10.47\,$\pm$\,2.29 & $<$\,45.0 &                                     2.04$_{-0.31}^{+0.63}$ & 8.13\,$\pm$\,0.04 & 2.6$_{-0.4}^{+0.2}$ & 24[38]               \\   
LESS79.2 & 1.98\,$\pm$\,0.40 & 40.90\,$\pm$\,7.00 & $<$\,14.86 & 17.72\,$\pm$\,3.74 & 20.78\,$\pm$\,3.52 & 719.92\,$\pm$\,15.00 &                         1.55$_{-0.18}^{+0.11}$ & 8.41\,$\pm$\,0.08 & 2.1$_{-0.0}^{+0.6}$ & 25[32]               \\   
LESS79.4 & 1.81\,$\pm$\,0.51 & $<$\,22.8 & $<$\,15.68 & $<$\,15.81 & $<$\,12.37 & $<$\,45.0 &                                                             4.60$_{-0.60}^{+1.20}$ & 8.46\,$\pm$\,0.11 & 1.9$_{-0.7}^{+2.6}$ & 51[38]               \\   
LESS80.1 & 4.03\,$\pm$\,0.86 & 48.80\,$\pm$\,7.00 & $<$\,13.14 & $<$\,13.13 & $<$\,9.14 & $<$\,45.0 &                                                     1.96$_{-0.14}^{+0.16}$ & 8.01\,$\pm$\,0.08 & 1.3$_{-0.3}^{+0.4}$ & 22[38]               \\   
LESS80.2 & 3.54\,$\pm$\,0.90 & $<$\,22.8 & $<$\,13.14 & $<$\,13.13 & $<$\,9.62 & 185.31\,$\pm$\,15.00 &                                                   1.37$_{-0.08}^{+0.17}$ & 8.72\,$\pm$\,0.10 & 0.4$_{-0.2}^{+0.5}$ & 18[19]               \\   
LESS82.1 & 1.93\,$\pm$\,0.47 & $<$\,22.8 & $<$\,12.31 & $<$\,8.80 & $<$\,7.00 & $<$\,45.0 &                                                               2.10$_{-0.44}^{+3.27}$ & 8.15\,$\pm$\,0.09 & 1.8$_{-1.0}^{+1.4}$ & 24[32]               \\   
LESS83.4 & 1.39\,$\pm$\,0.36 & $<$\,22.8 & $<$\,13.42 & $<$\,10.37 & $<$\,7.41 & $<$\,45.0 &                                                              0.57$_{-0.50}^{+1.54}$ & 7.81\,$\pm$\,0.10 & 0.3$_{-0.3}^{+0.6}$ & 13[19]               \\   

\end{tabular}

}
\end{table*}

\begin{table*}
{\tiny
\begin{tabular}{lcccccccccc}
\hline
\hline
ID          & S$_{870\mu m}$ &  S$_{\rm 1.4GHz}$ & S$_{500\mu m}$ & S$_{350\mu m}$ & S$_{250\mu m}$ & S$_{\rm 24\mu m}$     & $z_{\rm phot}$   & $\log$($M_{\rm dust}/M_{\odot}$) & $L_{\rm 8-1000\mu m}$ & $T_{\rm d}$ \\
            & (mJy)         & ($\mu$Jy)       & (mJy)        & (mJy)         &  (mJy)        & (mJy)               &                &                            & ($\times$\,10$^{12}$\,$L_{\odot}$) & (K) \\
\hline
LESS84.1 & 3.17\,$\pm$\,0.63 & 38.90\,$\pm$\,6.80 & 21.99\,$\pm$\,4.50 & 25.53\,$\pm$\,4.47 & 17.34\,$\pm$\,3.22 & 204.78\,$\pm$\,15.00 &               1.92$_{-0.07}^{+0.09}$ & 8.27\,$\pm$\,0.08 & 2.7$_{-0.7}^{+0.3}$ & 24[38]    \\     
LESS84.2 & 3.25\,$\pm$\,0.77 & $<$\,22.8 & $<$\,14.86 & $<$\,14.36 & 11.36\,$\pm$\,2.49 & 146.23\,$\pm$\,15.00 &                                        1.75$_{-0.19}^{+0.08}$ & 8.60\,$\pm$\,0.09 & 1.0$_{-0.5}^{+0.4}$ & 20[24]    \\     
LESS87.1 & 1.34\,$\pm$\,0.35 & 122.30\,$\pm$\,9.30 & $<$\,12.31 & 12.28\,$\pm$\,2.96 & 16.58\,$\pm$\,3.04 & 504.78\,$\pm$\,15.00 &                      3.20$_{-0.47}^{+0.08}$ & 8.78\,$\pm$\,0.10 & 3.2$_{-0.5}^{+0.2}$ & 41[32]    \\     
LESS87.3 & 2.44\,$\pm$\,0.59 & $<$\,22.8 & $<$\,10.60 & $<$\,9.15 & $<$\,7.00 & $<$\,45.0 &                                                             4.00$_{-0.30}^{+1.10}$ & 8.93\,$\pm$\,0.09 & 2.5$_{-1.0}^{+1.9}$ & 39[19]    \\     
LESS88.1 & 4.62\,$\pm$\,0.58 & 33.60\,$\pm$\,6.80 & 20.96\,$\pm$\,4.38 & 19.71\,$\pm$\,3.94 & 14.82\,$\pm$\,2.83 & 378.07\,$\pm$\,15.00 &               1.84$_{-0.11}^{+0.12}$ & 8.67\,$\pm$\,0.05 & 2.2$_{-0.8}^{+0.8}$ & 22[24]    \\     
LESS88.11 & 2.51\,$\pm$\,0.71 & $<$\,22.8 & 22.44\,$\pm$\,4.57 & 19.37\,$\pm$\,3.92 & 11.71\,$\pm$\,2.62 & $<$\,45.0 &                                  2.57$_{-0.12}^{+0.04}$ & 8.29\,$\pm$\,0.11 & 2.3$_{-0.6}^{+2.2}$ & 27[38]    \\    
LESS88.2 & 2.14\,$\pm$\,0.50 & $<$\,22.8 & $<$\,16.18 & $<$\,16.47 & $<$\,15.50 & $<$\,45.0 &                                                           5.20$_{-1.20}^{+0.60}$ & 8.62\,$\pm$\,0.09 & 2.3$_{-1.0}^{+3.1}$ & 60[37]    \\     
LESS88.5 & 2.86\,$\pm$\,0.72 & 38.20\,$\pm$\,6.60 & 12.56\,$\pm$\,3.08 & 22.98\,$\pm$\,4.26 & 17.05\,$\pm$\,3.13 & 143.40\,$\pm$\,15.00 &               2.30$_{-0.50}^{+0.11}$ & 8.29\,$\pm$\,0.10 & 4.5$_{-0.9}^{+1.2}$ & 31[41]    \\     
LESS92.2 & 2.42\,$\pm$\,0.68 & $<$\,22.8 & $<$\,11.79 & $<$\,11.76 & $<$\,7.00 & $<$\,45.0 &                                                            1.90$_{-0.75}^{+0.28}$ & 9.01\,$\pm$\,0.11 & 0.4$_{-0.2}^{+0.2}$ & 22[19]    \\     
LESS94.1 & 3.18\,$\pm$\,0.52 & $<$\,22.8 & 10.87\,$\pm$\,2.76 & 11.44\,$\pm$\,2.85 & 8.19\,$\pm$\,1.92 & 168.21\,$\pm$\,15.00 &                         2.87$_{-0.64}^{+0.37}$ & 8.94\,$\pm$\,0.07 & 1.9$_{-0.6}^{+2.0}$ & 30[24]    \\     
LESS98.1 & 4.78\,$\pm$\,0.60 & 145.00\,$\pm$\,8.20 & 36.86\,$\pm$\,6.33 & 65.47\,$\pm$\,6.65 & 64.74\,$\pm$\,6.76 & 217.63\,$\pm$\,15.00 &              1.63$_{-0.09}^{+0.17}$ & 8.67\,$\pm$\,0.05 & 4.7$_{-0.0}^{+0.7}$ & 27[35]    \\     
LESS99.1 & 2.05\,$\pm$\,0.43 & $<$\,22.8 & $<$\,10.60 & $<$\,8.00 & $<$\,7.00 & $<$\,45.0 &                                                             5.00$_{-0.60}^{+1.20}$ & 8.59\,$\pm$\,0.08 & 2.2$_{-0.8}^{+2.1}$ & 48[32]    \\     
LESS102.1 & 3.08\,$\pm$\,0.50 & 38.50\,$\pm$\,9.10 & 10.65\,$\pm$\,2.70 & 12.79\,$\pm$\,3.03 & 9.16\,$\pm$\,2.08 & 202.01\,$\pm$\,15.00 &               1.76$_{-0.18}^{+0.16}$ & 8.56\,$\pm$\,0.06 & 1.4$_{-0.6}^{+0.8}$ & 22[24]    \\    
LESS103.3 & 1.43\,$\pm$\,0.41 & $<$\,22.8 & $<$\,14.86 & $<$\,12.68 & $<$\,10.14 & $<$\,45.0 &                                                          4.40$_{-0.70}^{+0.70}$ & 8.82\,$\pm$\,0.11 & 1.6$_{-0.7}^{+1.8}$ & 49[19]    \\    
LESS107.1 & 1.91\,$\pm$\,0.39 & $<$\,22.8 & $<$\,12.58 & 10.24\,$\pm$\,2.67 & 10.77\,$\pm$\,2.32 & $<$\,45.0 &                                          3.75$_{-0.08}^{+0.09}$ & 8.28\,$\pm$\,0.08 & 3.5$_{-0.8}^{+1.1}$ & 40[41]    \\    
LESS107.3 & 1.46\,$\pm$\,0.40 & $<$\,22.8 & $<$\,12.04 & $<$\,13.97 & $<$\,12.94 & $<$\,45.0 &                                                          2.12$_{-0.81}^{+1.54}$ & 7.88\,$\pm$\,0.10 & 1.1$_{-0.9}^{+2.5}$ & 31[38]    \\    
LESS110.1 & 4.11\,$\pm$\,0.47 & $<$\,22.8 & $<$\,10.60 & 11.96\,$\pm$\,2.89 & 10.41\,$\pm$\,2.33 & $<$\,45.0 &                                          2.55$_{-0.50}^{+0.70}$ & 8.28\,$\pm$\,0.05 & 5.2$_{-1.0}^{+0.7}$ & 27[40]    \\    
LESS110.5 & 2.39\,$\pm$\,0.60 & $<$\,22.8 & $<$\,10.60 & $<$\,9.94 & $<$\,9.14 & $<$\,45.0 &                                                            3.70$_{-1.20}^{+0.40}$ & 8.94\,$\pm$\,0.10 & 2.3$_{-1.3}^{+2.0}$ & 39[19]    \\    
LESS112.1 & 7.62\,$\pm$\,0.49 & $<$\,22.8 & 20.82\,$\pm$\,4.33 & 20.86\,$\pm$\,4.02 & 18.13\,$\pm$\,3.20 & 161.26\,$\pm$\,15.00 &                       1.95$_{-0.26}^{+0.15}$ & 9.36\,$\pm$\,0.03 & 1.0$_{-0.0}^{+0.2}$ & 23[19]    \\    
LESS114.1 & 2.99\,$\pm$\,0.78 & $<$\,22.8 & 18.20\,$\pm$\,4.00 & 17.95\,$\pm$\,3.71 & 16.34\,$\pm$\,2.97 & $<$\,45.0 &                                  3.00$_{-0.50}^{+1.40}$ & 8.86\,$\pm$\,0.10 & 6.2$_{-3.9}^{+1.3}$ & 34[19]    \\    
LESS114.2 & 1.98\,$\pm$\,0.50 & 97.80\,$\pm$\,6.80 & 10.87\,$\pm$\,2.80 & 32.58\,$\pm$\,5.07 & 41.62\,$\pm$\,5.37 & 513.43\,$\pm$\,15.00 &              1.56$_{-0.07}^{+0.07}$ & 8.64\,$\pm$\,0.10 & 2.6$_{-0.3}^{+1.0}$ & 33[32]    \\    
LESS116.1 & 3.08\,$\pm$\,0.47 & $<$\,22.8 & 18.41\,$\pm$\,3.98 & 17.53\,$\pm$\,3.65 & 11.89\,$\pm$\,2.43 & $<$\,45.0 &                                  3.54$_{-0.87}^{+1.47}$ & 8.51\,$\pm$\,0.06 & 2.6$_{-0.2}^{+3.5}$ & 38[38]   \\    
LESS116.2 & 3.42\,$\pm$\,0.57 & 41.90\,$\pm$\,6.80 & 17.26\,$\pm$\,3.83 & 17.09\,$\pm$\,3.60 & 8.67\,$\pm$\,2.00 & $<$\,45.0 &                          4.02$_{-2.19}^{+1.19}$ & 8.80\,$\pm$\,0.07 & 4.2$_{-0.7}^{+1.0}$ & 39[33]   \\    
LESS118.1 & 3.20\,$\pm$\,0.54 & 43.50\,$\pm$\,7.80 & 12.30\,$\pm$\,3.00 & 16.71\,$\pm$\,3.56 & 14.43\,$\pm$\,2.79 & $<$\,45.0 &                         2.26$_{-0.23}^{+0.50}$ & 8.24\,$\pm$\,0.07 & 2.4$_{-0.3}^{+0.7}$ & 29[38]   \\    
LESS119.1 & 8.27\,$\pm$\,0.54 & $<$\,22.8 & $<$\,12.85 & $<$\,11.29 & $<$\,7.65 & $<$\,45.0 &                                                           3.50$_{-0.35}^{+0.95}$ & 8.95\,$\pm$\,0.03 & 4.3$_{-2.6}^{+3.7}$ & 27[24]   \\    
LESS122.1 & 3.69\,$\pm$\,0.42 & 207.50\,$\pm$\,7.40 & 32.08\,$\pm$\,5.79 & 42.54\,$\pm$\,5.70 & 48.26\,$\pm$\,5.73 & 1479.99\,$\pm$\,15.00 &            2.06$_{-0.06}^{+0.05}$ & 8.90\,$\pm$\,0.05 & 6.3$_{-0.5}^{+0.4}$ & 32[33]   \\    
LESS124.1 & 3.64\,$\pm$\,0.57 & 27.70\,$\pm$\,7.80 & 19.83\,$\pm$\,4.19 & 19.25\,$\pm$\,3.84 & 9.16\,$\pm$\,2.09 & 80.41\,$\pm$\,15.00 &                6.07$_{-1.16}^{+0.94}$ & 8.69\,$\pm$\,0.06 & 7.2$_{-2.5}^{+1.1}$ & 54[41]   \\    
LESS124.4 & 2.24\,$\pm$\,0.58 & $<$\,22.8 & $<$\,13.42 & $<$\,13.56 & 7.31\,$\pm$\,1.78 & $<$\,45.0 &                                                   5.60$_{-1.20}^{+0.60}$ & 8.56\,$\pm$\,0.10 & 4.3$_{-1.3}^{+1.3}$ & 48[35]   \\    
LESS126.1 & 2.23\,$\pm$\,0.55 & 25.10\,$\pm$\,6.70 & $<$\,10.60 & $<$\,8.00 & 7.02\,$\pm$\,1.75 & 219.61\,$\pm$\,15.00 &                                1.82$_{-0.08}^{+0.28}$ & 8.29\,$\pm$\,0.10 & 1.3$_{-0.7}^{+0.5}$ & 19[24]   \\     

\hline
\end{tabular}

\footnotesize{Notes: 
  3$\sigma$ upper limits are given in the case of non-detections.  $^{*}$ We provide two estimates of the dust temperature; the first value corresponds to the characteristic dust temperature as measured from a modified black-body fit to the far-infrared photometry at the photometric redshift.  The second value (in parenthesis) corresponds to the wavelength of the peak of the best-fit dust SED template and assuming $\lambda_{\rm peak}T_{\rm d}$\,=\,2.897\,$\times$\,10$^{-3}$\,m.K.}
\label{table:phot}
}
\end{table*}

\begin{figure*}
  \centerline{\psfig{file=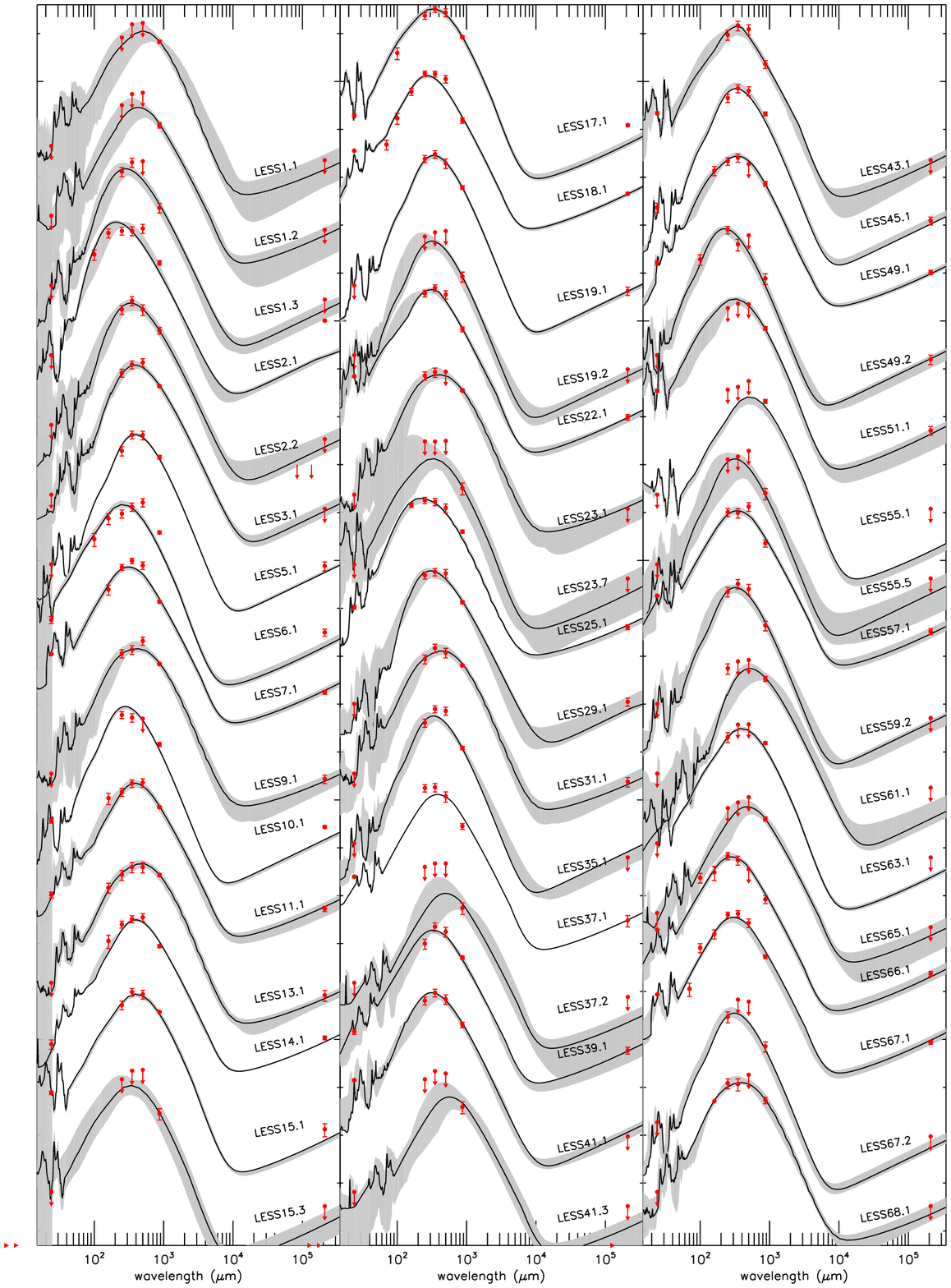,angle=0,width=6.5in}}
\caption{SEDs of the ALMA SMGs in our sample.  In each case,
  the SPIRE photometry has been deblended.  The solid black
  curve shows the best-fit SED, and the grey region shows the range of
  acceptable solutions.
}
\label{fig:allSEDs}
\end{figure*}

\begin{figure*}
  \centerline{\psfig{file=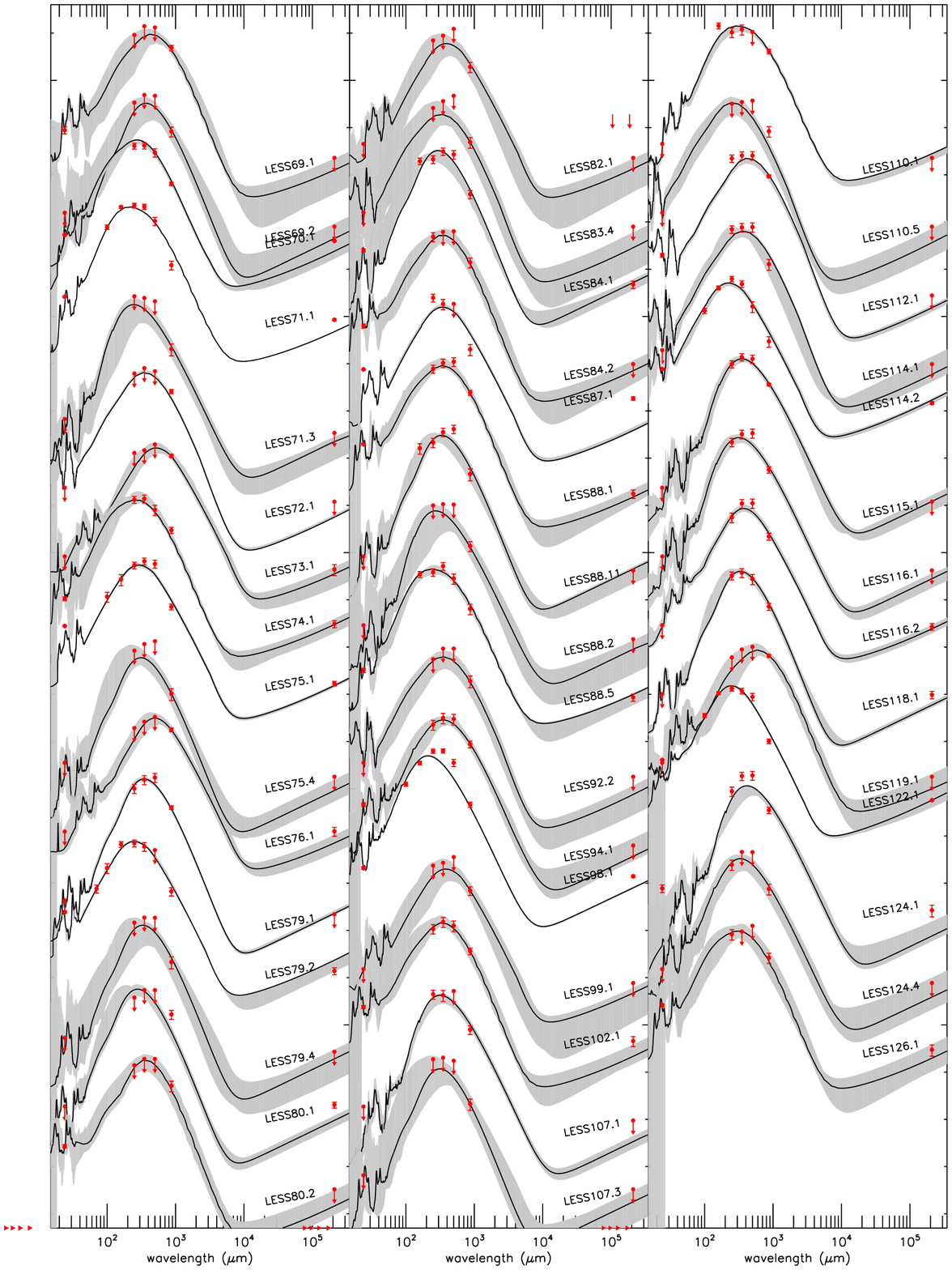,angle=0,width=6.5in}}
\end{figure*}

\end{document}